\documentclass[12pt]{elsarticle} 
\usepackage{amsmath}
\usepackage{textcomp}
\usepackage{hyperref}
\usepackage{graphicx}
\usepackage{subfigure}
\begin{document}

\title{Dynamics of episodic transient correlations in currency exchange rate returns and their predictability}


\author{Milan \v{Z}ukovi\v{c}$^{1,2}$}
\ead{milan.zukovic@upjs.sk}

\address{
$^1$ SORS Research a.s., Moyzesova 38, 040 01 Ko\v{s}ice, Slovakia \\
$^2$ Department of Theoretical Physics and Astrophysics, Faculty of Science, P. J. \v{S}af\'arik University, 
     Park Angelinum 9, 041 54 Ko\v{s}ice, Slovakia
          }

\begin{abstract}
We study the dynamics of the linear and non-linear serial dependencies in financial time series in a rolling window framework. In particular, we focus on the detection of episodes of statistically significant two- and three-point correlations in the returns of several leading currency exchange rates that could offer some potential for their predictability. We employ a rolling window approach in order to capture the correlation dynamics for different window lengths and analyze the distributions of periods with statistically significant correlations. We find that for sufficiently large window lengths these distributions fit well to power-law behavior. We also measure the predictability itself by a hit rate, i.e. the rate of consistency between the signs of the actual returns and their predictions, obtained from a simple correlation-based predictor. It is found that during these relatively brief periods the returns are predictable to a certain degree and the predictability depends on the selection of the window length.
\end{abstract}



\maketitle


\section{Introduction}
\label{intro}
Recent empirical studies suggested that returns of financial time series depart from the random walk hypothesis by showing a certain degree of long-term or short-term dependent relationships, thus violating the weak-form efficient market hypothesis~\cite{lo,eom1,eom2}. The efficiency/non-efficiency can, for example, be assessed by evaluation of the Hurst exponent~\cite{hur} or the approximate entropy~\cite{pin}, as measures of long-term memory and randomness in time series. Since the efficient market hypothesis in its weakest form implies that the returns should be serially uncorrelated, its validity can also be assessed by looking for evidence of significant linear~\cite{fam,gra} and non-linear~\cite{hin-pat1} serial autocorrelations (henceforth, correlations) in the returns. A systematic review of literature on the weak-form market efficiency of stock markets using a wide array of statistical tests, such as the linear serial correlations, unit root, low-dimensional chaos, nonlinear serial dependence and long memory was provided by Lim and Brooks~\cite{lim11}. However, the market efficiency vary with time and the character of possible serial dependencies is not known. Moreover, they are of transient nature, showing up in random intervals just for a short time, which makes it difficult to exploit them for prediction purposes~\cite{ser}.  

In several previous studies~\cite{ser,bro2,bon1,bon2} the non-linearities were investigated by calculation of the bicorrelation test statistic, due to Hinich~\cite{hin}, and performing the windowed-test procedure~\cite{hin-pat2,hin-pat3}. In order to capture the time variation, the data were split into a set of non-overlapping windows, the length of which was set to some ad-hoc fixed value. However, considering the episodic and transient nature of the correlations, these may or may not be detected, depending on the window length used. Furthermore, the window length also influences the onset and offset of the significant correlations. Shorter lengths facilitate quicker response to changes in the correlation strength and can help pinpoint the arrival and disappearance of the transient dependences, but on the other hand, they may lack adequate statistical power. 

In order to track the evolution of market efficiency over time, a rolling-window approach has been adopted to calculate a time-varying Hurst exponent~\cite{caj1,caj2,alv,wan}. Lim~\cite{lim07} has proposed a similar rolling window approach for computing the bicorrelation test for various stock markets~\cite{lim1,lim2}. It was shown that the market efficiency is not a static property that remains unchanged throughout the entire estimation period. Thus, examination of the presence of significant correlations in the rolling sample framework can provide a useful tool for tracking the changing degree of weak-form marker efficiency over time as well as for its ranking among different markets. In the currency exchange market, Brooks and Hinich~\cite{bro1} found that Sterling exchange rates are characterized by transient epochs of dependencies surrounded by long periods of white noise.

The objective of this paper is to study the dynamics of the linear and non-linear serial dependencies in financial time series, using a portmanteau test procedure in a rolling window framework. In particular, we focus on detection of episodes of statistically significant two- and three-point correlations in the returns of several leading currency exchange rates that could offer some potential for their predictability. We employ a rolling window approach in order to capture the correlation dynamics for different window lengths. In comparison with similar previous studies, our new contribution is using tools from statistical physics in order to analyze distributions of the periods with statistically significant correlations and relate them to the predictability obtained from a simple correlation-based predictor.


\section{Data}
\label{data}

\begin{figure}[!ht]
  \begin{center}
    \subfigure[USD/EUR]{\label{fig:rates_pair1}
    \includegraphics[scale=0.38]{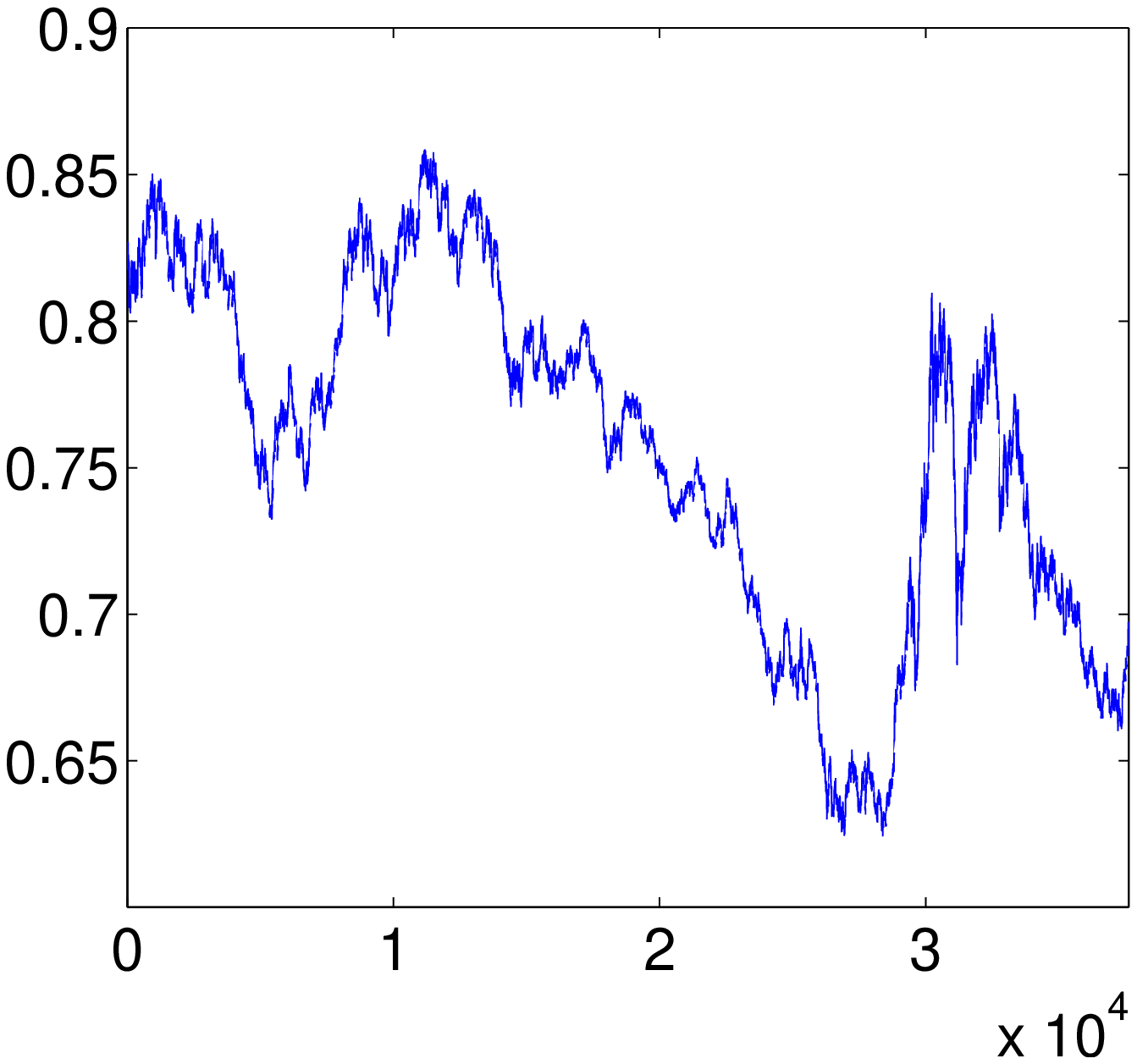}}
    \subfigure[CAD/EUR]{\label{fig:rates_pair2}
    \includegraphics[scale=0.38]{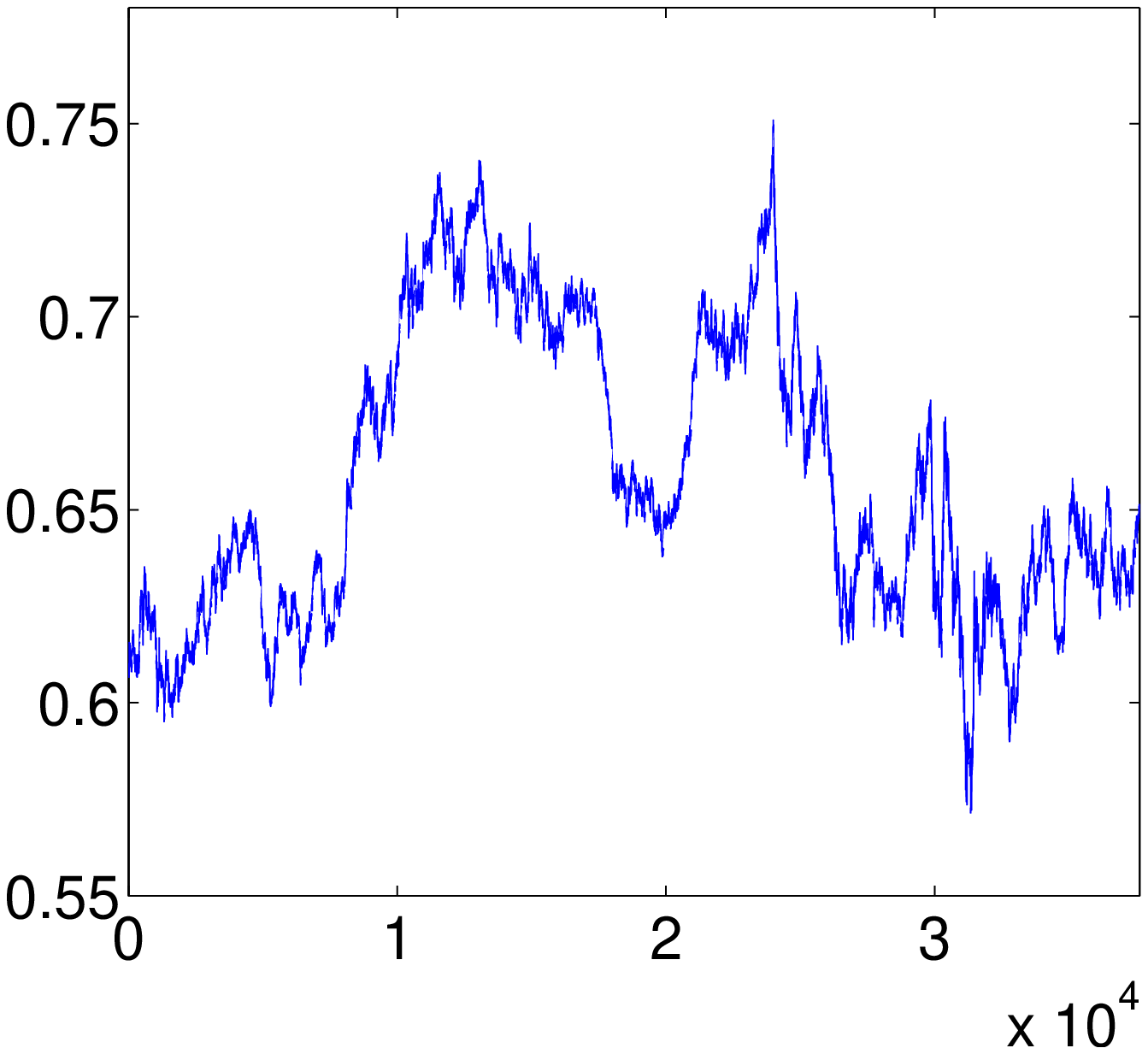}}
    \subfigure[CHF/EUR]{\label{fig:rates_pair3}
    \includegraphics[scale=0.38]{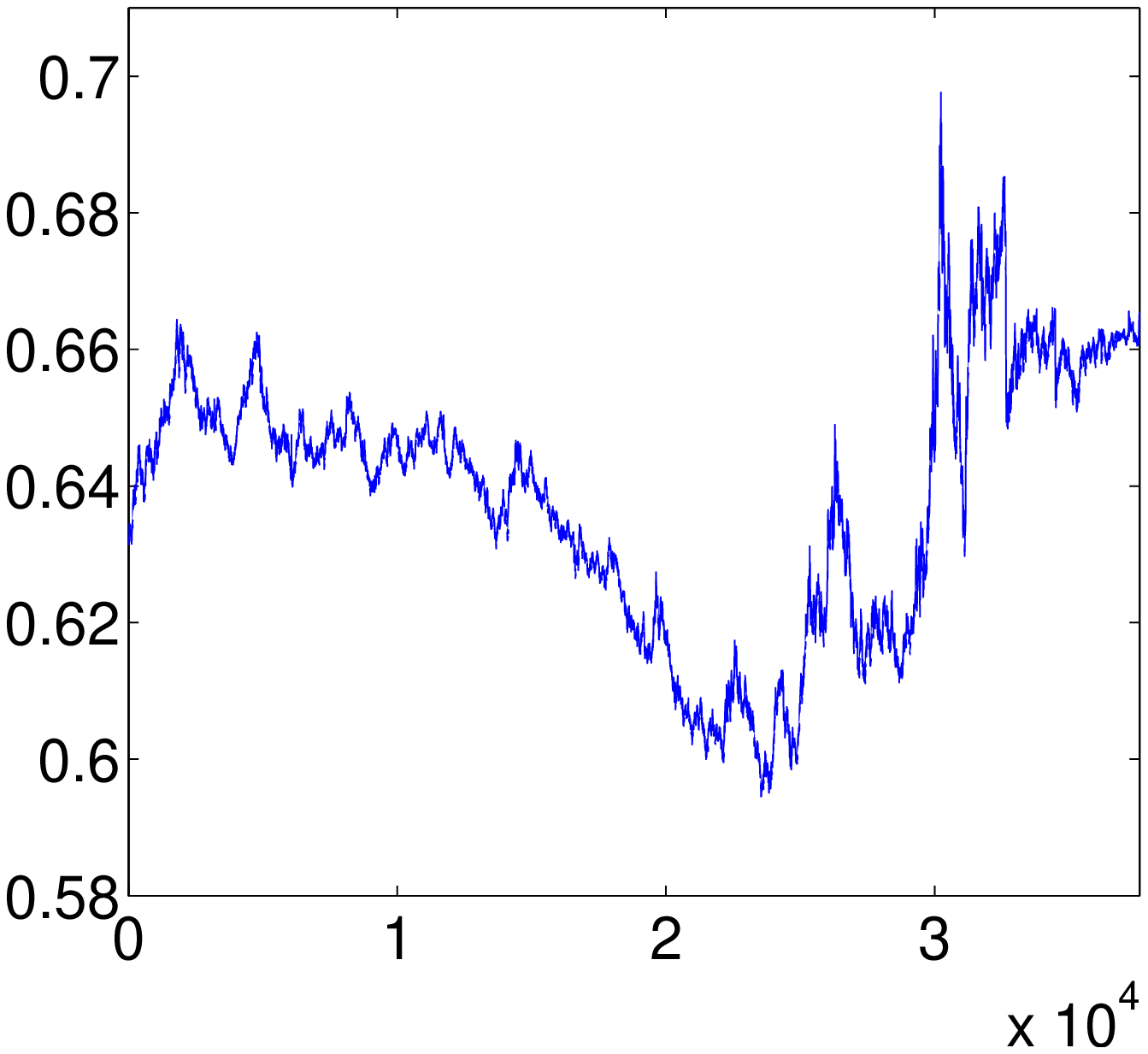}}
    \subfigure[GBP/EUR]{\label{fig:rates_pair4}
    \includegraphics[scale=0.38]{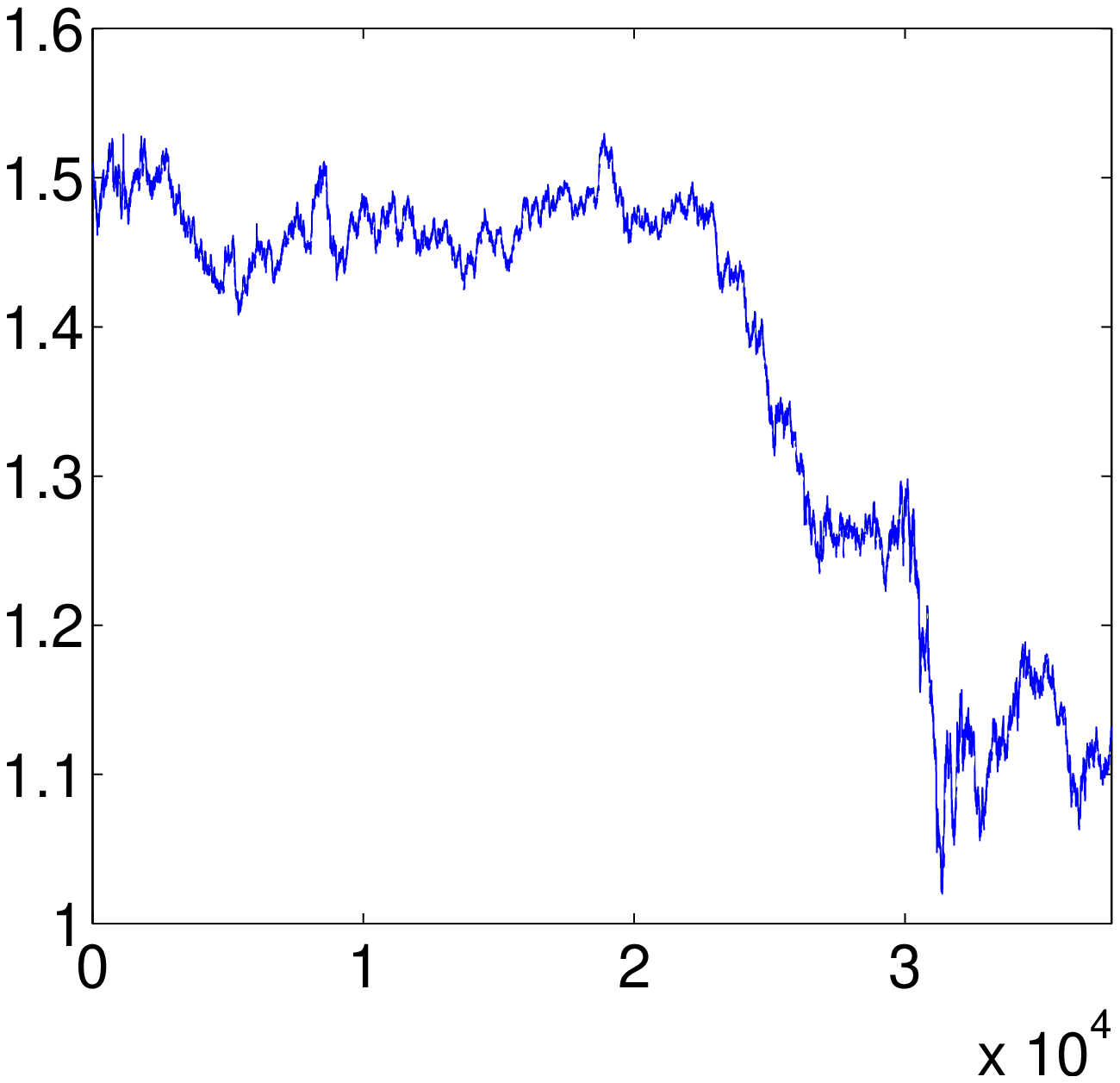}}
    \subfigure[JPY/EUR]{\label{fig:rates_pair5}
    \includegraphics[scale=0.38]{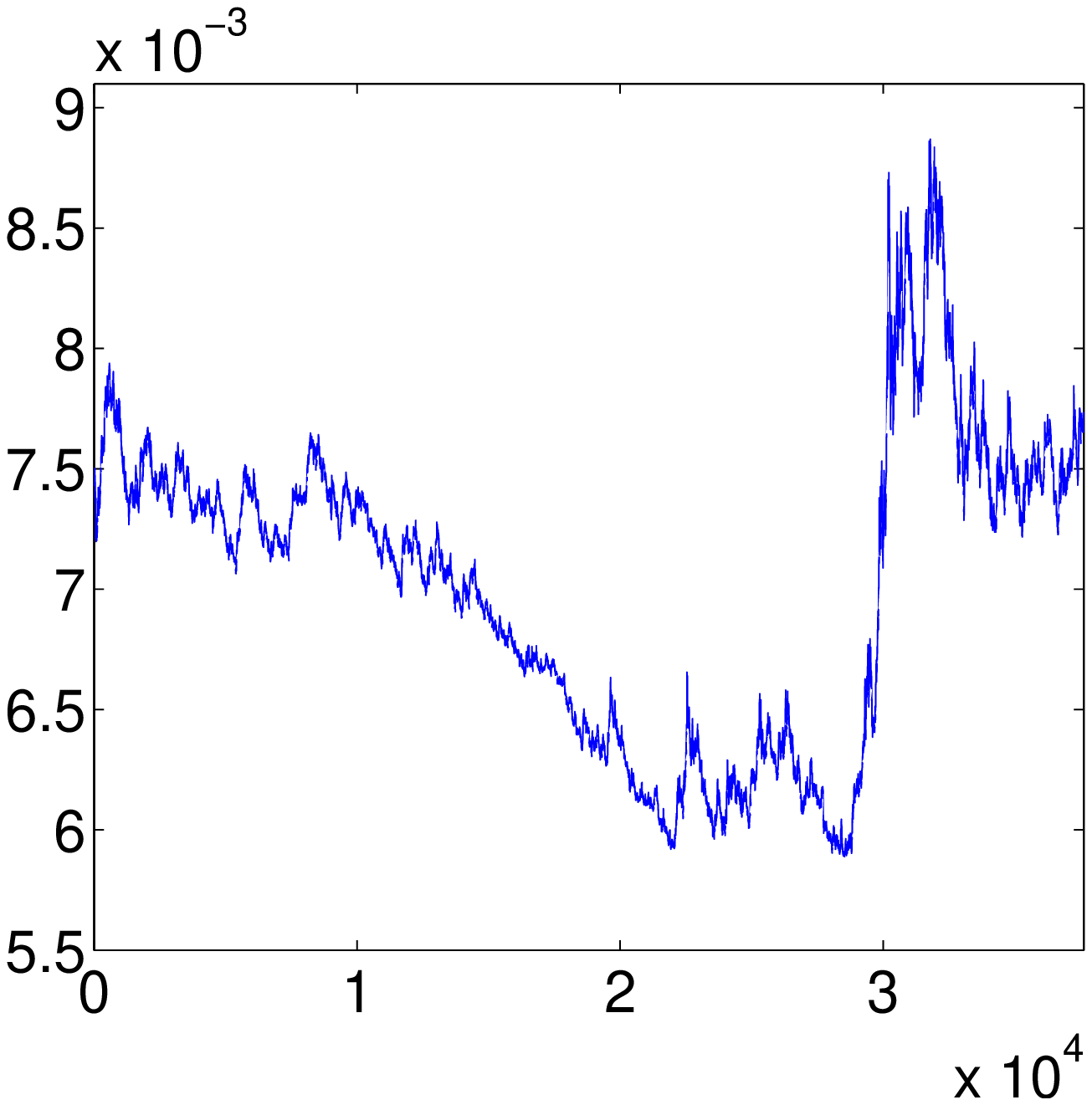}}
  \end{center}
\caption{Nominal exchange rates.}
  \label{fig:rates}
\end{figure}

\begin{figure}[!ht]
  \begin{center}
  \subfigure[USD/EUR]{\label{fig:logret_pair1}
    \includegraphics[scale=0.38]{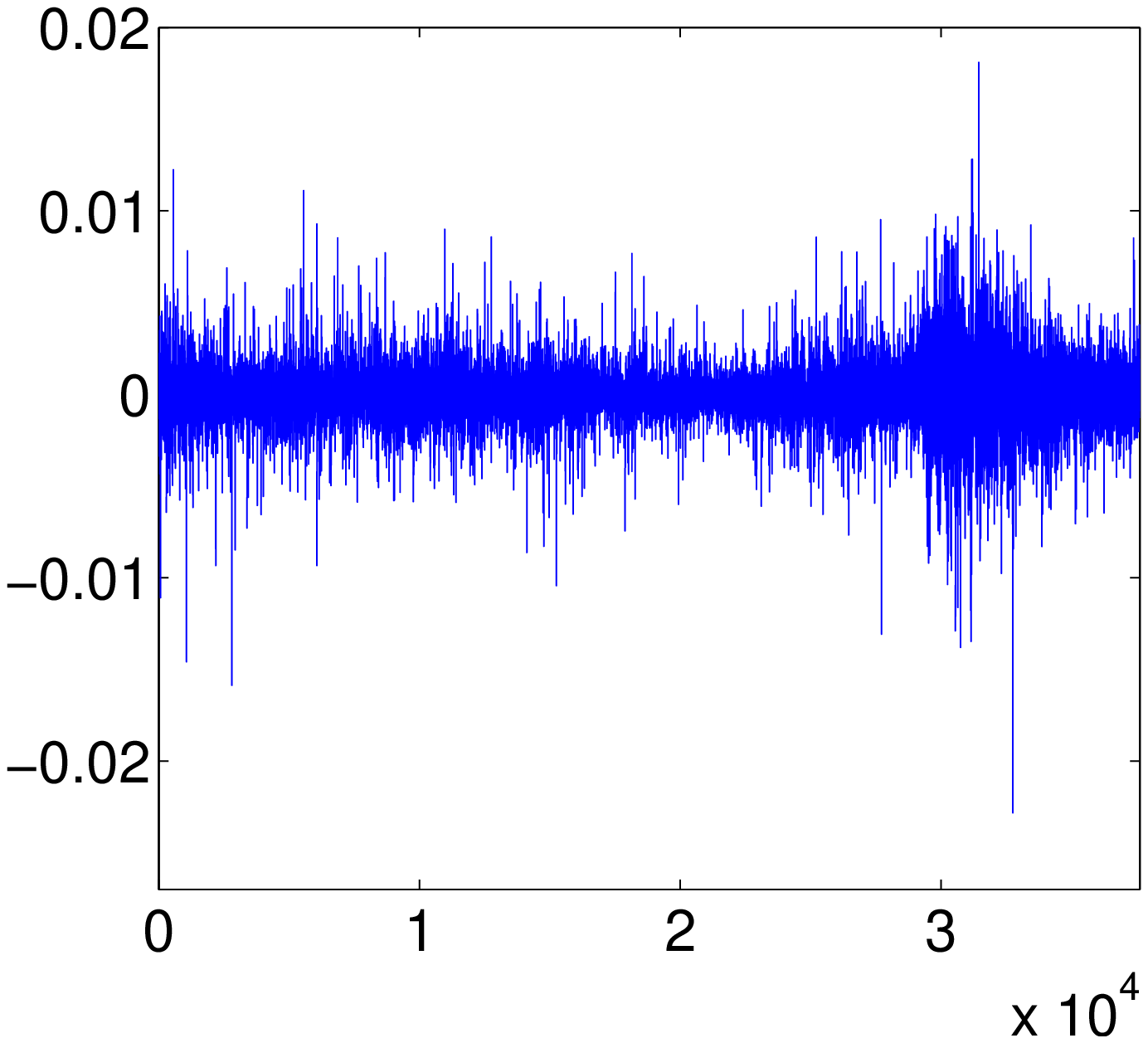}}
    \subfigure[CAD/EUR]{\label{fig:logret_pair2}
    \includegraphics[scale=0.38]{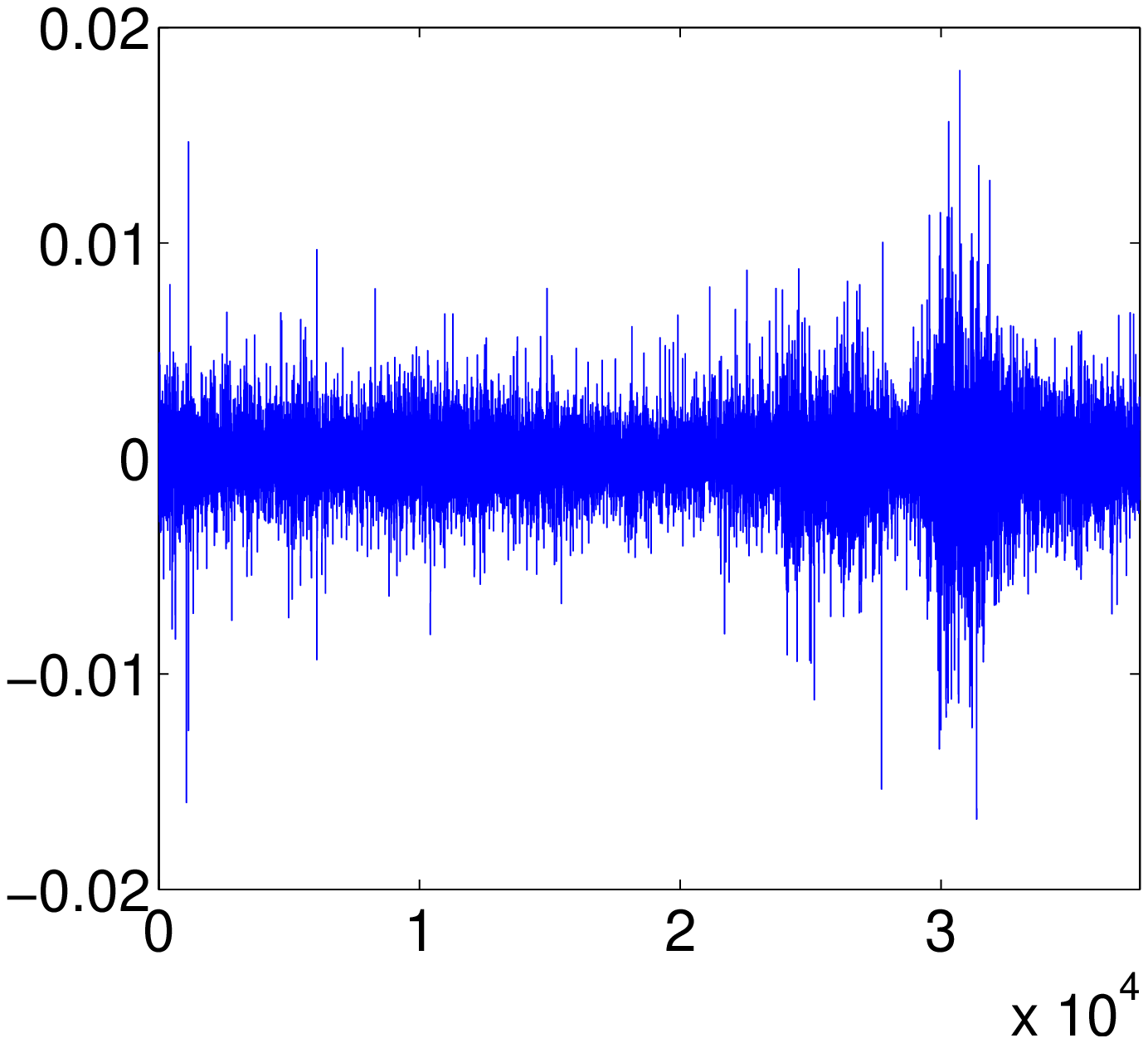}}
    \subfigure[CHF/EUR]{\label{fig:logret_pair3}
    \includegraphics[scale=0.38]{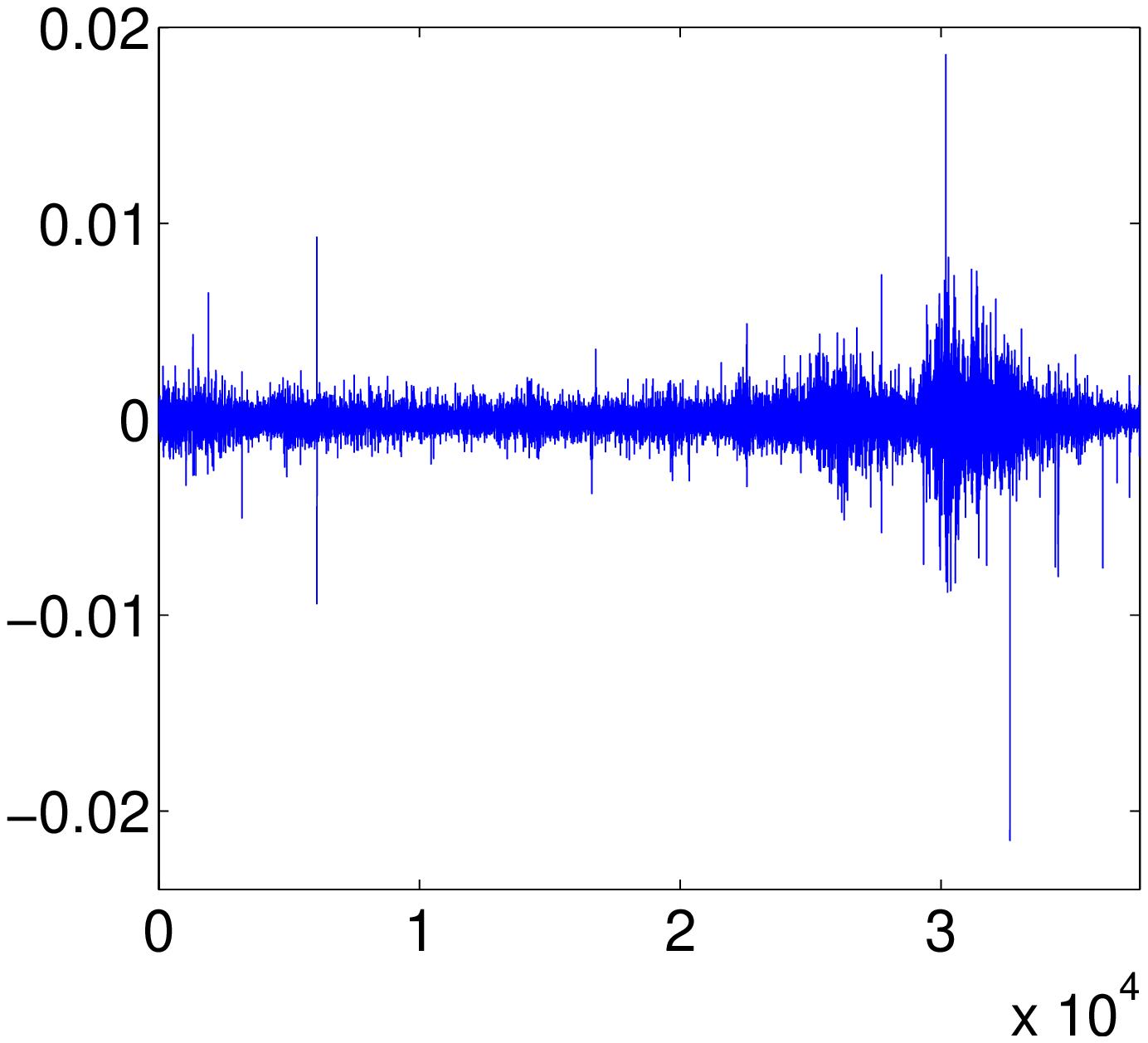}}
    \subfigure[GBP/EUR]{\label{fig:logret_pair4}
    \includegraphics[scale=0.38]{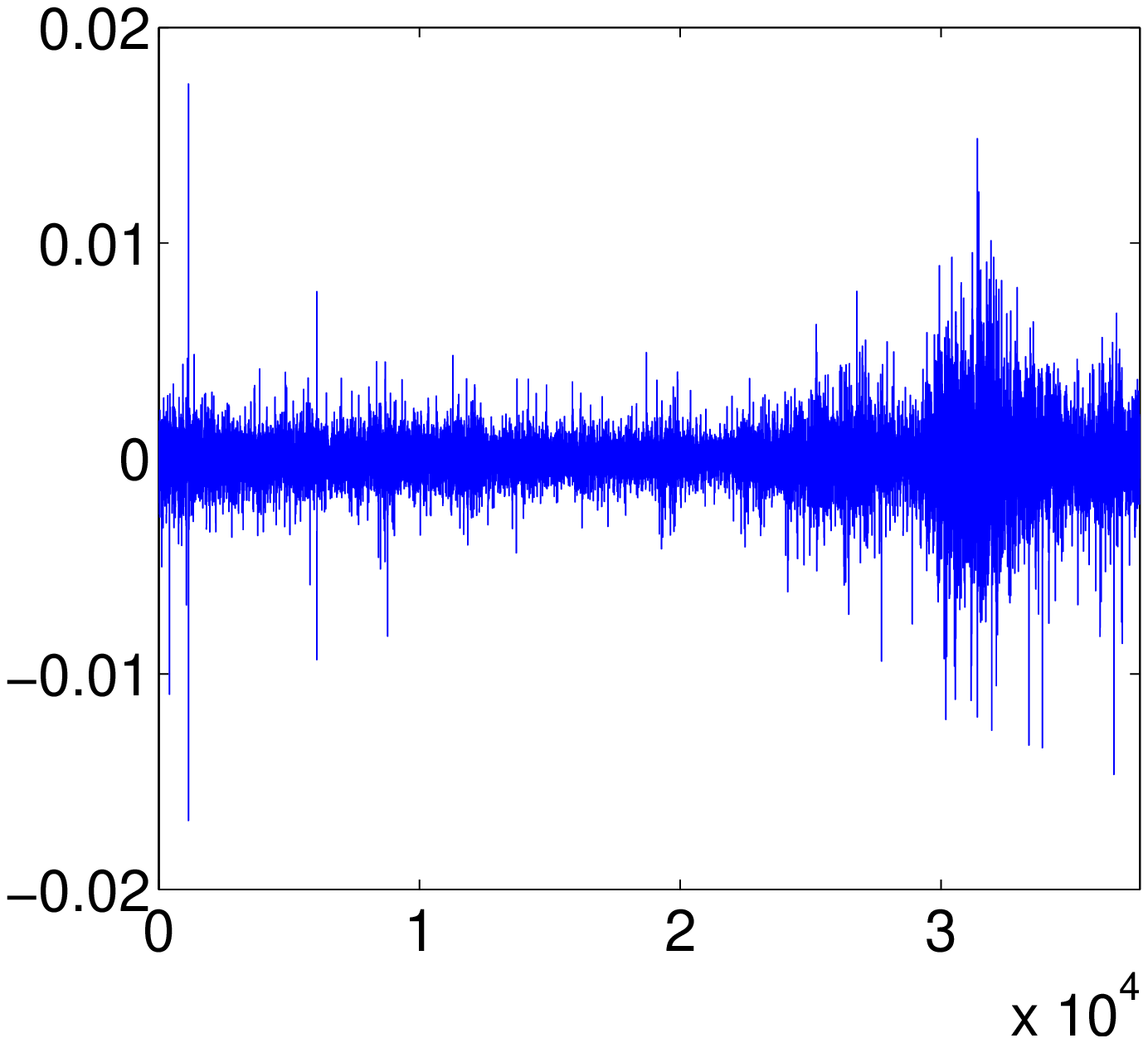}}
    \subfigure[JPY/EUR]{\label{fig:logret_pair5}
    \includegraphics[scale=0.38]{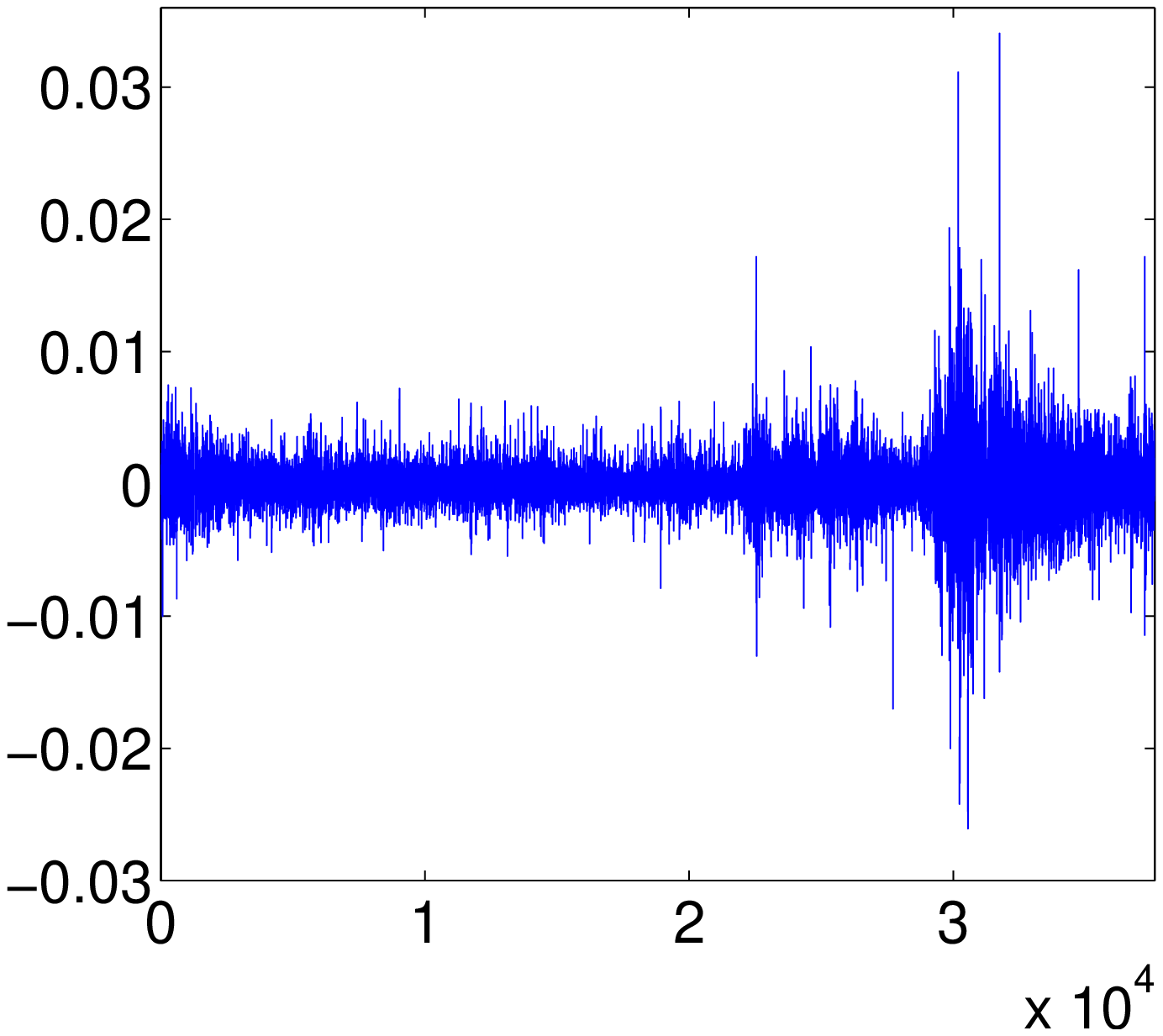}}
  \end{center}
\caption{Logarithmic returns.}
  \label{fig:logret}
\end{figure}

\begin{figure}[!ht]
  \begin{center}
    \subfigure[USD/EUR]{\label{fig:hist_logret_pair1}
    \includegraphics[scale=0.38]{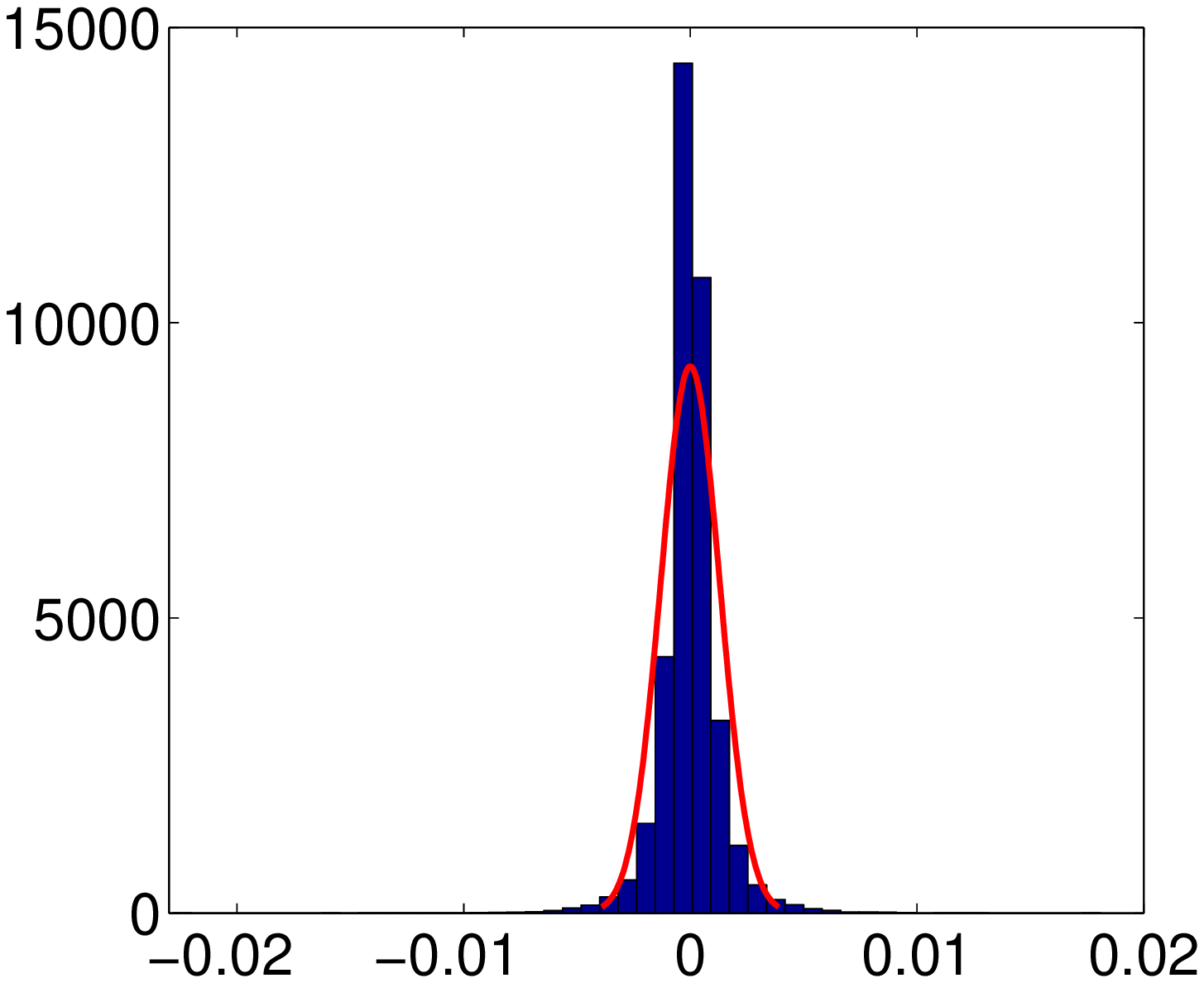}}
    \subfigure[CAD/EUR]{\label{fig:hist_logret_pair2}
    \includegraphics[scale=0.38]{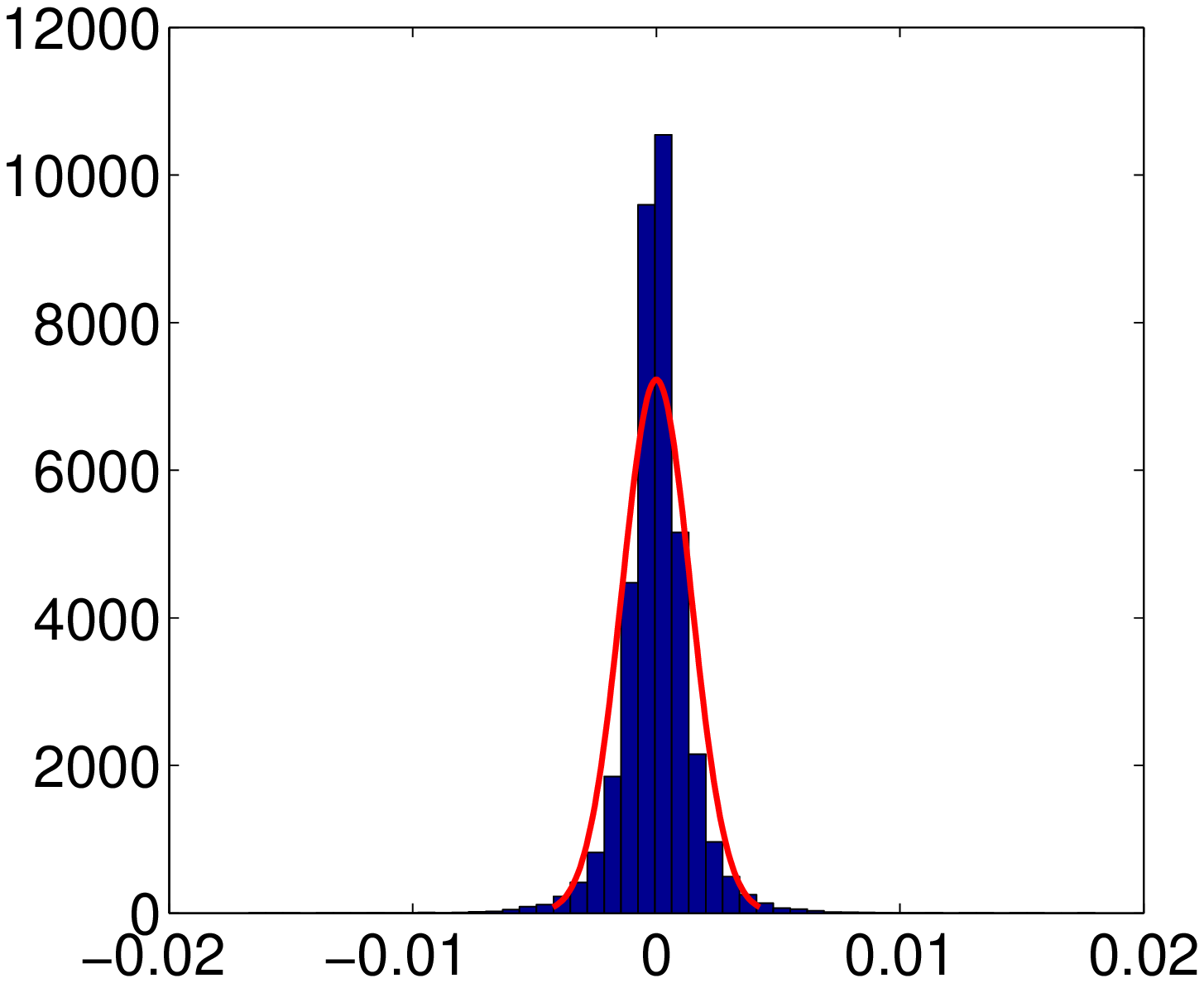}}
    \subfigure[CHF/EUR]{\label{fig:hist_logret_pair3}
    \includegraphics[scale=0.38]{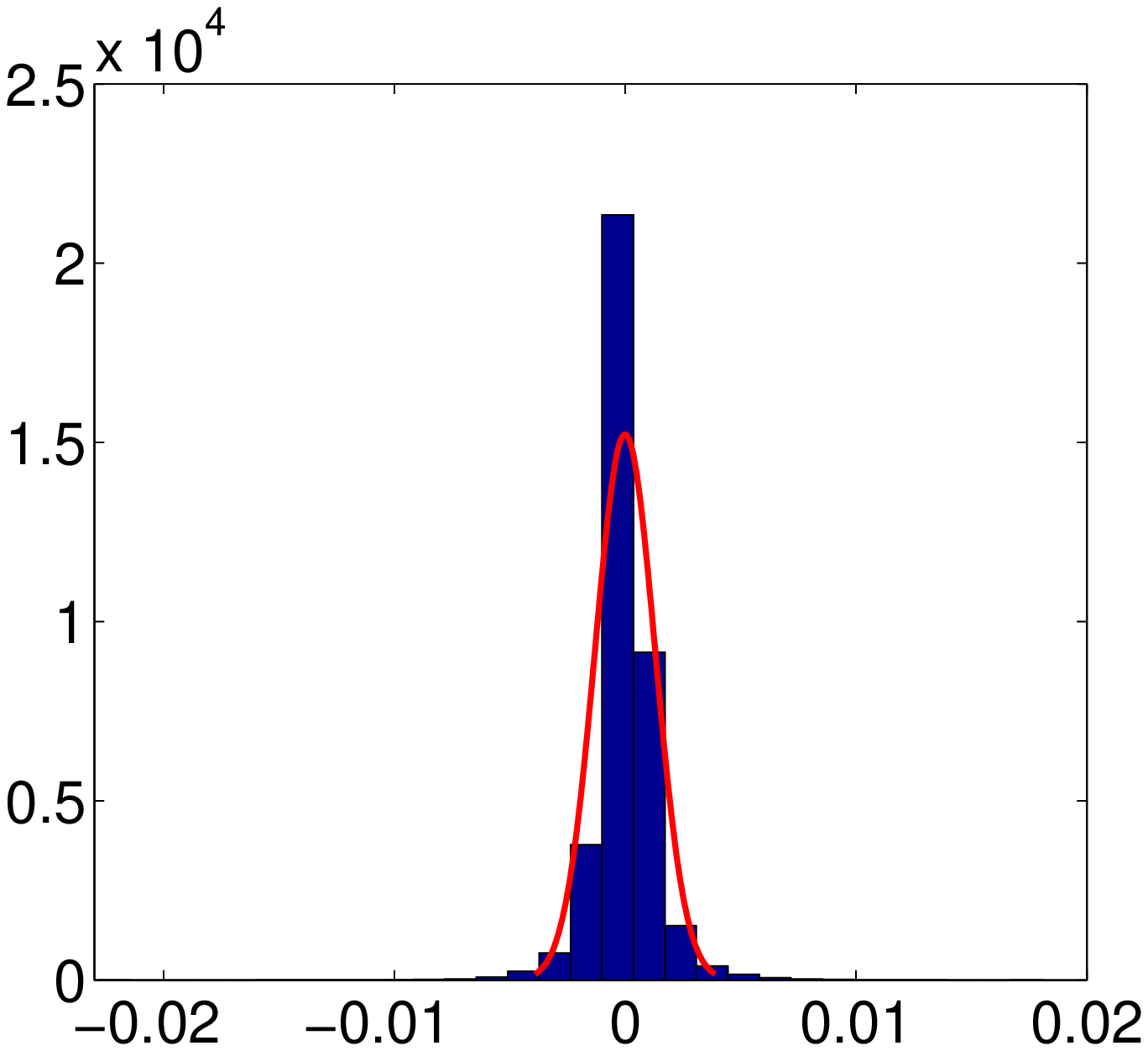}}
    \subfigure[GBP/EUR]{\label{fig:hist_logret_pair4}
    \includegraphics[scale=0.38]{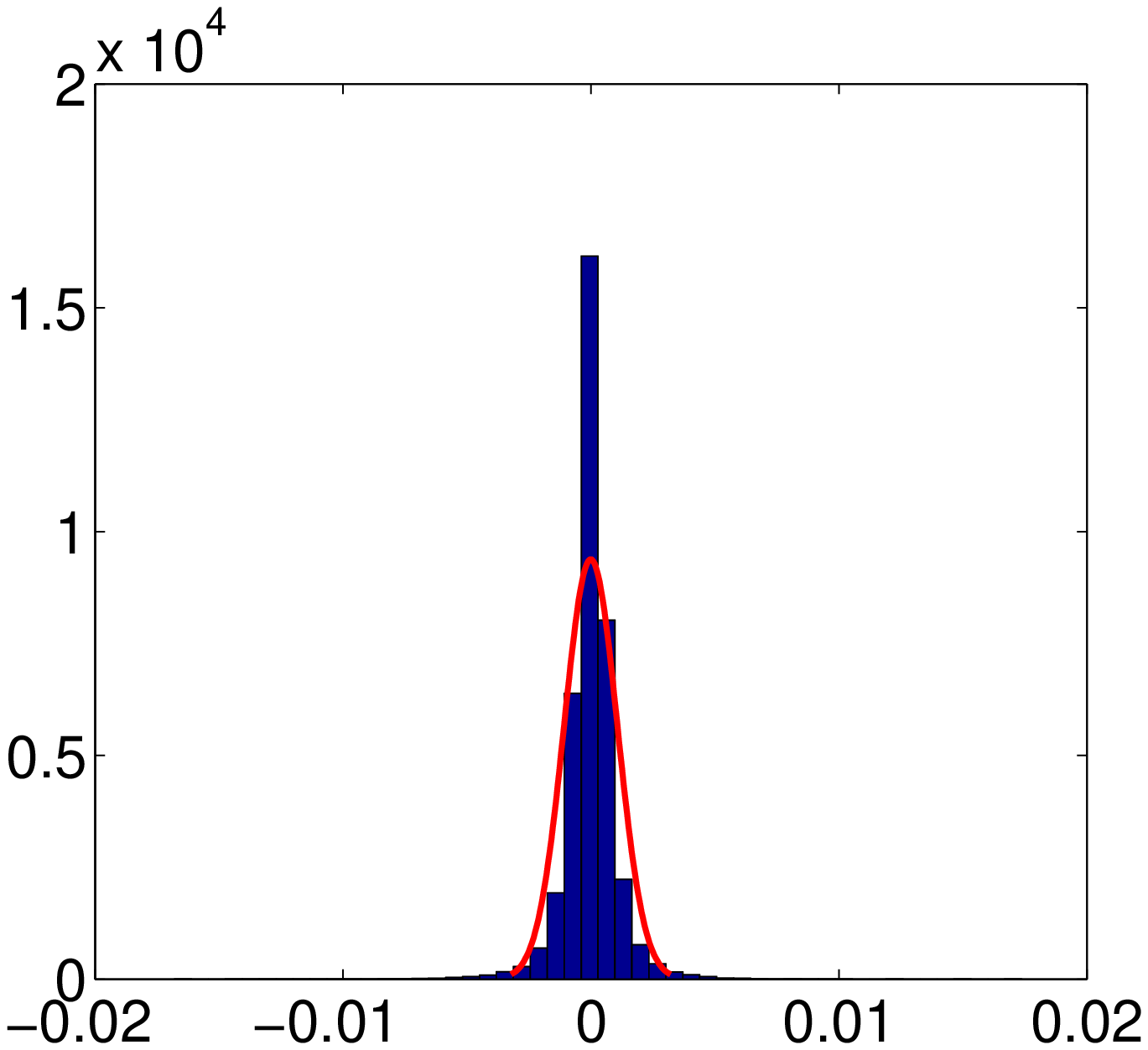}}
    \subfigure[JPY/EUR]{\label{fig:hist_logret_pair5}
    \includegraphics[scale=0.38]{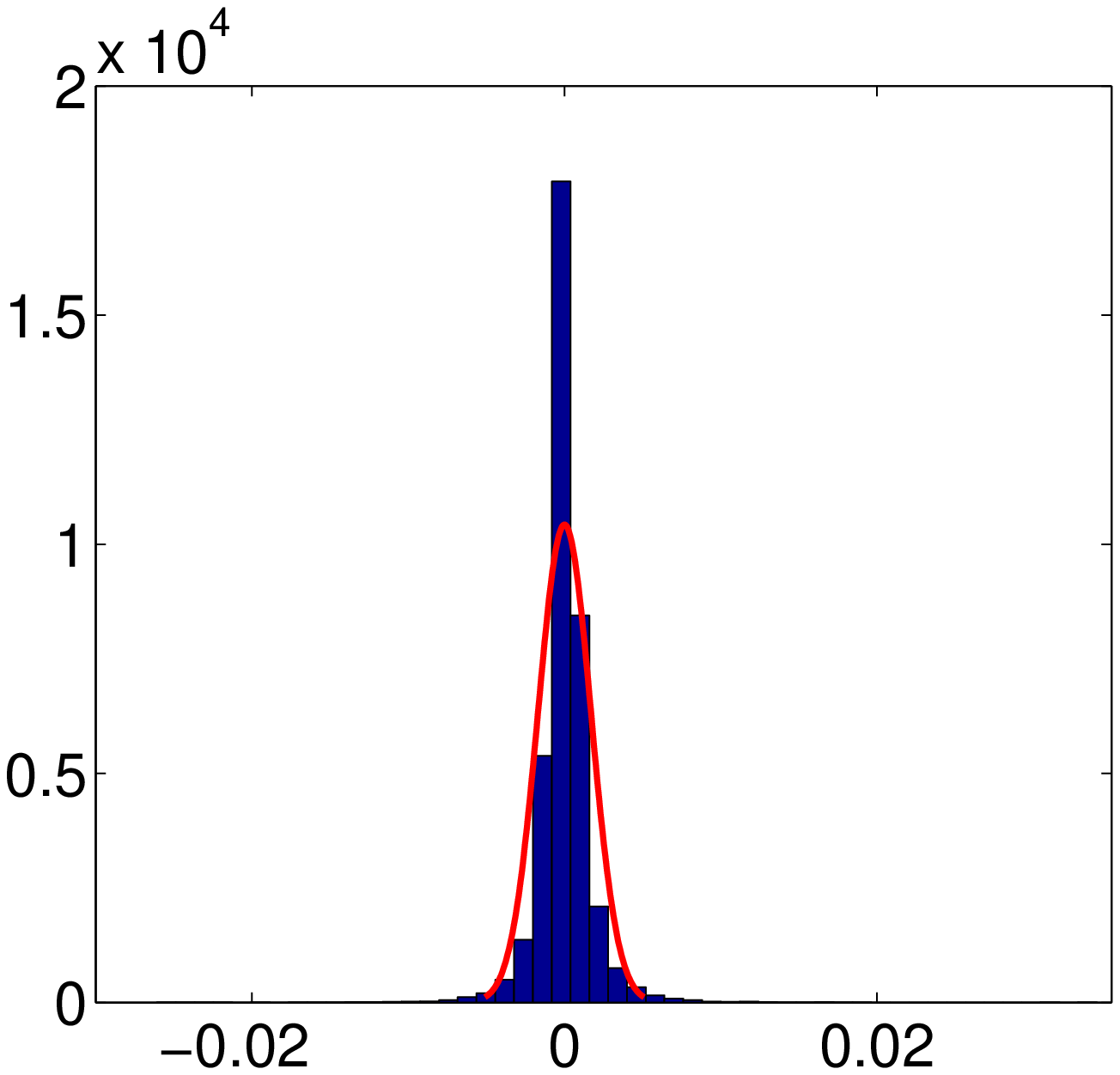}}
  \end{center}
\caption{Frequency distributions of the logarithmic returns with the Gaussian curve fit.}
  \label{fig:hist_logret}
\end{figure}

We use hourly average exchange rates per Euro (EUR) on the US dollar (USD), Canadian dollar (CAD), Swiss franc (CHF), British pound (GBP) and Japanese yen (JPY). The sample period is from March 1, 2004 to December 17, 2009 (a total of 37,620 observations). In Fig.~\ref{fig:rates} we plot the nominal exchange rates of the respective currencies vs the Euro. In Figs.~\ref{fig:logret} and~\ref{fig:hist_logret} we show first logarithmic differences of the nominal exchange rates and their frequency distributions, respectively. The histograms and the summary statistics, presented in Table~\ref{tab:SummaryStatistics}, indicate that the data deviate from the normal distribution. This observation is also supported by the Jarque-Bera~\cite{jb} normality tests, which reject the null hypothesis of normality with very low $p$-values. 

\vspace{15mm}

\begin{table}
	\centering
		\begin{tabular}{lccccc}
		\hline
    & USD/EUR  & CAD/EUR & CHF/EUR & GBP/EUR & JPY/EUR \\
		\hline \hline
			Mean & -4.266e-06 & 1.715e-06 & 1.389e-06 & -7.607e-06 & 1.042e-06 \\
			Median & 0.00000 & 0.00000 & 0.00000 & 0.00000 & -1.796e-05 \\
			Maximum & 0.0181 & 0.0180  & 0.0186  & 0.0174  &  0.0341 \\
			Minimum & -0.0228 & -0.0167 & -0.0215 & -0.0168 &  -0.0261 \\
			Std deviation & 1.3e-03 & 1.4e-03 & 0.7e-03 & 1.1e-03 & 1.7e-03 \\
			Skewness & -0.1521 & -0.0486 & -0.3913 & -0.3651 & 0.2721 \\
			Kurtosis & 17.59 & 13.14 & 52.95 & 22.03 & 27.41 \\
			JB normality test statistic & 333,840 & 161,300 & 3,912,000 & 568,380 & 934,710 \\
			$p$-value &  $<$ 1.0e-03 &  $<$ 1.0e-03 &  $<$ 1.0e-03 & $<$ 1.0e-03 & $<$ 1.0e-03 \\
			\hline
		\end{tabular}
	\caption{Summary statistics of logarithmic returns}
	\label{tab:SummaryStatistics}
\end{table}

\section{Methods}
\label{methods}
\subsection{Detection of significant correlations and bicorrelations in a rolling window}
Let $R(t)$ represent the logarithmic returns of the length $N$, defined by $R(t)=\ln (P(t+1)/P(t))$, where $P(t)$ are the nominal exchange rates and $t$ is a time label, $t=1,...,N$. Within the rolling window framework we test for both linear and non-linear correlations in a window of a specified length $n$, that is rolling one point forward eliminating the first observation and including the next one, until the last observation. The data in each window are standardized to have zero mean and unit standard deviation, as follows: 
\begin{equation}
\label{stand} x(t)=\frac{R(t)-M_R}{S_R},
\end{equation}
\noindent where $M_R$ and $S_R$ are the sample mean and standard deviation of the window. The window-test procedure~\cite{hin-pat2,hin-pat3} uses correlation and bicorrelation potmanteu tests for the detection of linear and non-linear serial dependencies within a time window. The null hypothesis is that the standardized data $x(t)$ in each window are realizations of a stationary pure white noise process. The alternative hypothesis is that the process has some non-zero correlations $C_{xx}(r)={\rm E}[x(t)x(t+r)]$ or bicorrelations $C_{xxx}(r,s)={\rm E}[x(t)x(t+r)x(t+s)]$ within $0<r<s \le L<n$, where $L$ is the number of lags. The sample correlations and bicorrelations in a rolling window are calculated as follows: 
\begin{equation}
\label{corr} C_{xx}(r)=\frac{1}{(n-r)}\sum_{t=1}^{n-r}x(t)x(t+r), 
\end{equation}
\begin{equation}
\label{bicorr} C_{xxx}(r,s)=\frac{1}{(n-s)}\sum_{t=1}^{n-s}x(t)x(t+r)x(t+s).
\end{equation}
The test statistics in a rolling window for non-zero correlations $H_{xx}$ and bicorrelations $H_{xxx}$, are respectively given by:
\begin{equation}
\label{corr_p} H_{xx}=\sum_{r=1}^{L}(n-r)C_{xx}^2(r) \sim \chi^{2}(L)
\end{equation}
and
\begin{equation}
\label{bicorr_p} H_{xxx}=\sum_{s=2}^{L}\sum_{r=1}^{s-1}(n-s)C_{xxx}^2(r,s) \sim \chi^{2}((L-1)L/2).
\end{equation}
We consider several window lengths $n=2^k,\ k=3,...,10.$ The number of lags is specified as $L=n^b$, where
$0<b<0.5$ is a user-specified parameter. In order to account only for the presumably most relevant lags, we 
chose a fixed value of $L=2$ (i.e. $r=1,s=2$), which, for example, for $n=2^4$ corresponds to $b=0.25$, as 
used in Ref.~\cite{ser}.

We note that in order to identify non-linearities by the $H_{xxx}$ test statistics, data 
pre-whitening is necessary. This is achieved by filtration of the linear 
component by an autoregressive AR(p) fit. The p order of the AR(p)
model is chosen between 1 and $n-1$ as the optimizer of Schwarz's Bayesian Criterion~\cite{lut,neu}. 
The null hypothesis of linear/non-linear correlation is accepted or rejected in each 
window at a risk level of 5\%.

\subsection{Simple correlation-based predictor}
Above we identify episodes with significant correlations which indicate transient inefficiency of the market. Emergence of such episodes offers some opportunity for the prediction of the price change. In the following we define a simple predictor of the price change direction in the intervals with significant correlations. Prediction of the price change direction $\hat{I}$ ($+1$ - increase, $-1$ - decrease, 0 - no prediction) at the time step $t+1$ is based on the assumption of consistency between the signs of the neighboring price changes and the sample correlation in a window:
\begin{equation}
\label{predict} \hat{I}(t+1)={\rm sign}(C_{xx}(1)x(t))\theta(\alpha - p_{xx}),
\end{equation}
\noindent where $p_{xx}$ is a $p$-value of the $H_{xx}$ statistic, $\theta$ is the step function ($\theta=1$ if $p_{xx} < \alpha$ and 0 otherwise), and $\alpha$ determines the confidence level for no correlation null hypothesis rejection (we used $\alpha=0.05$). Hence, if the sample correlation in a given time window with significant correlation is positive (negative) the predictor predicts the same (opposite) price change direction as the last observed change. The predictor's performance is evaluated by a hit rate, i.e. the rate of consistency between the signs of the actual returns and their predictions. We note that we avoided using a more complicated prediction model that would involve parameters. Instead, we tried to link the presence of significant correlations with the degree of predictability as directly as possible, thus eliminating the influence of the parameters' inference on the prediction. By using the current simple model the only relevant issue is weather the magnitude of the estimated correlation (positive or negative) significantly differs from zero or not.

\section{Results}
\label{results}
\subsection{Frequency and cluster size distribution of significant $H_{xx}$ and $H_{xxx}$ windows}
\begin{figure}[!t]
  \begin{center}
  \subfigure{\label{fig:percHxx}
    \includegraphics[scale=.49]{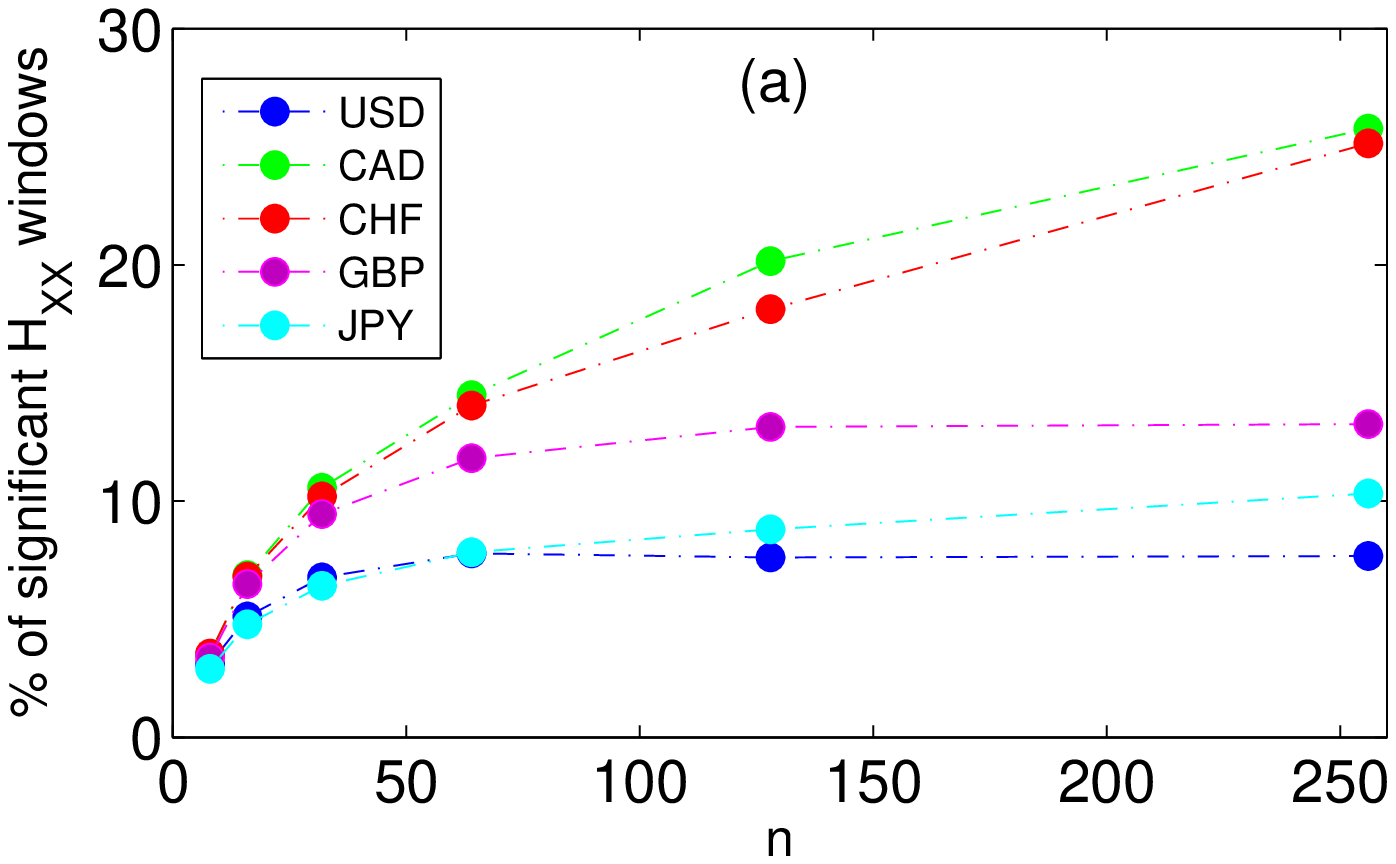}}
      \subfigure{\label{fig:percHxxx}
    \includegraphics[scale=.49]{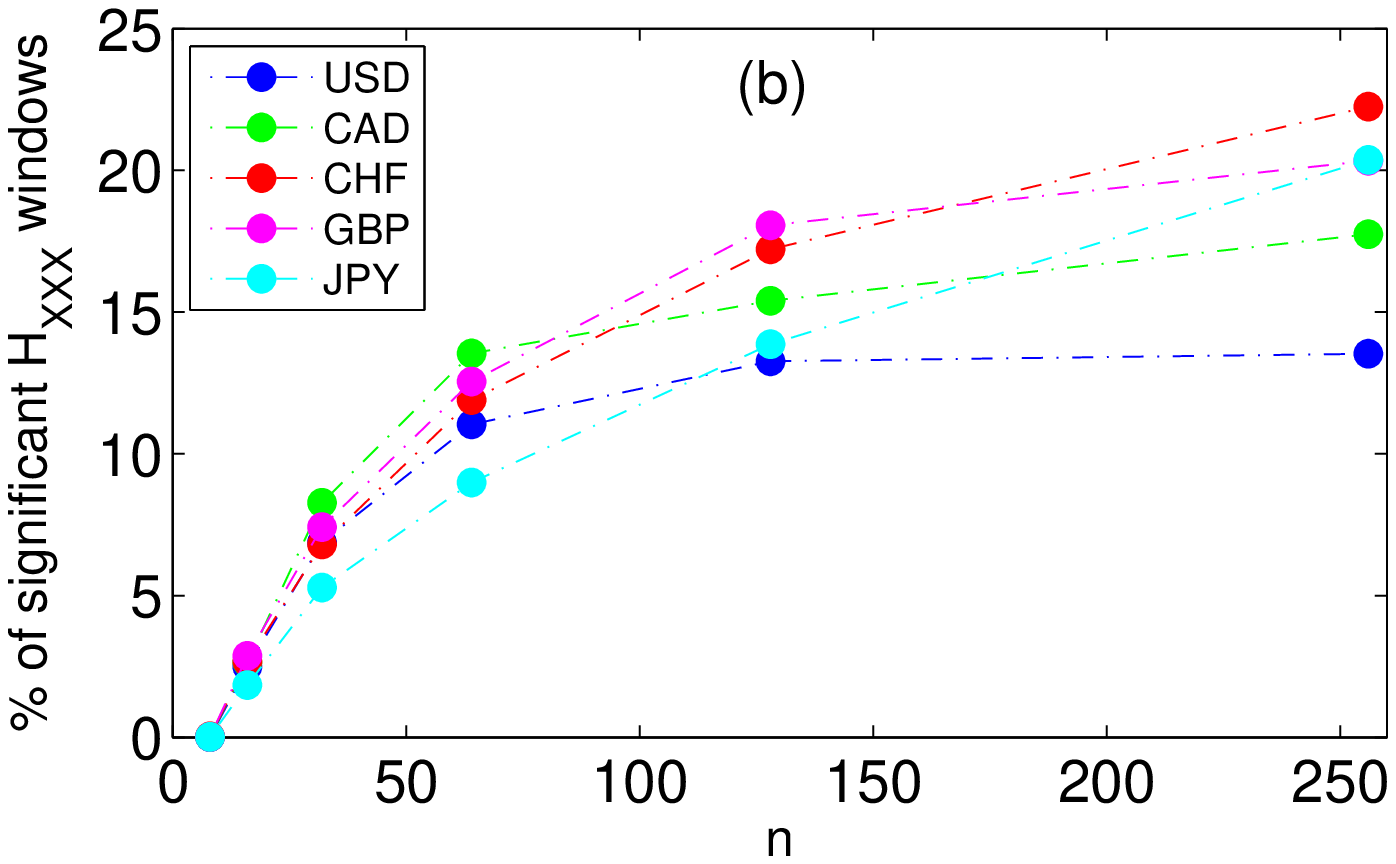}}
  \end{center}
  \caption{Percentage of the significant (a) $H_{xx}$ and (b) $H_{xxx}$ rolling windows for different currency pairs as a function of the window length $n$.}
  \label{fig:perc}
\end{figure}

The arrival of periods with significant correlations and their lengths are stochastic processes and their 
evaluation can shed some light on the overall efficiency of a particular market as well as the interpretation 
of the events, such as crises, government regulations, etc., that could have lead to the increase or decrease of 
the efficiency~\cite{wan,lim1,lim2}. Lim et al.~\cite{lim1,lim2} proposed the percentage of time windows in which the market departs from efficiency as an indicator for assessing the relative efficiency. In Fig.~\ref{fig:perc} we plot the percentage of the significant (a) $H_{xx}$ and (b) $H_{xxx}$ rolling windows for different currency pairs as a function of the window length $n$. We can observe that the percentage of the significant $H_{xx}$ and $H_{xxx}$ windows tends to increase with increasing $n$, with some currency pairs (e.g., CHF/EUR) being more and some less (e.g., USD/EUR) sensitive to the variation of $n$. Based on the efficiency indicator proposed by Lim et al.~\cite{lim1,lim2}, the largest (smallest) efficiency is found in USD/EUR (CHF/EUR) pair, featuring up to 13.5 (22.2) \% of the significant $H_{xxx}$ windows for the considered values of $n$.

\begin{figure}[!t]
  \begin{center}
  \subfigure{\label{fig:log-log_pair2a}
    \includegraphics[scale=.49]{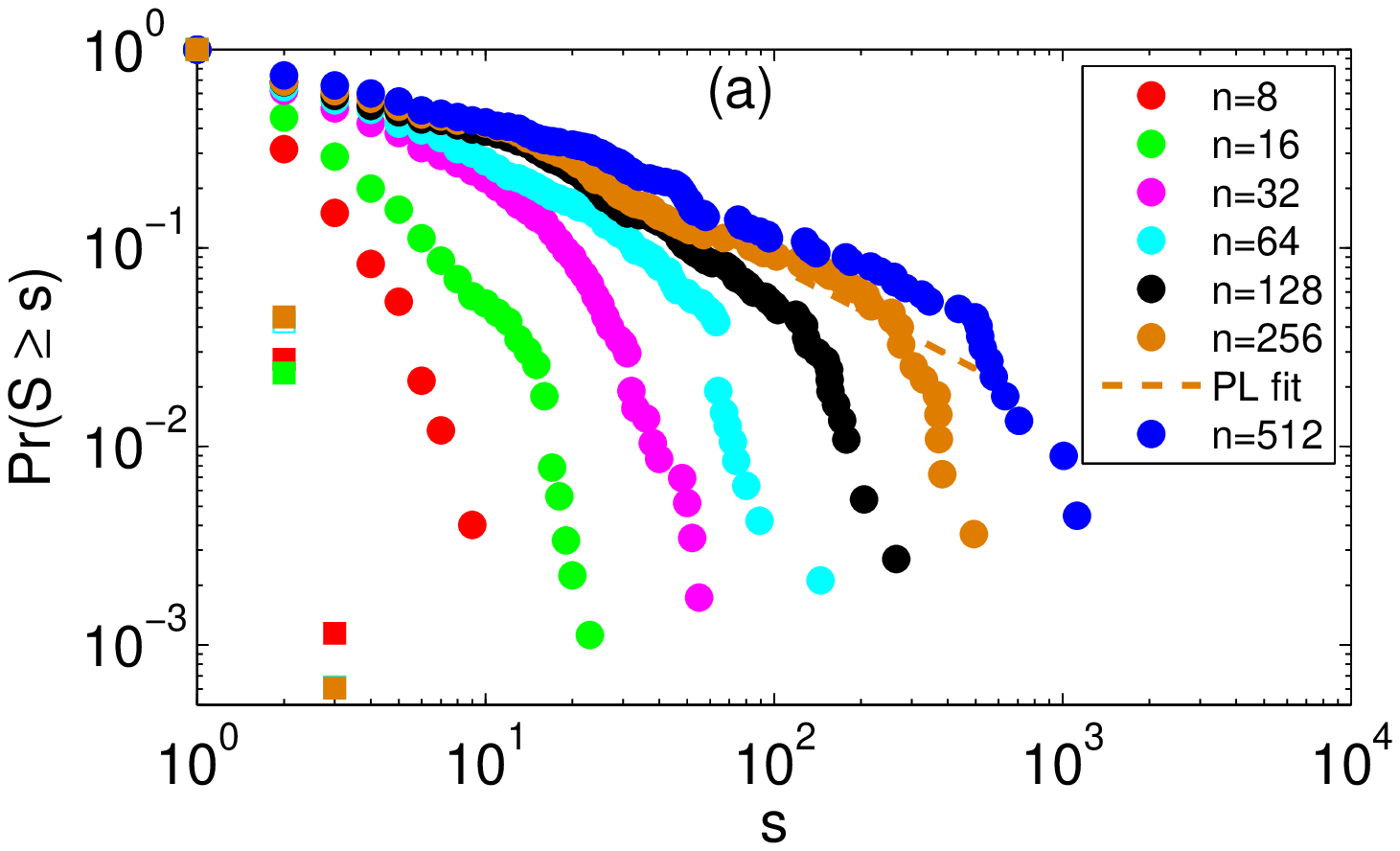}}
    \subfigure{\label{fig:log-log_pair2b}
    \includegraphics[scale=.49]{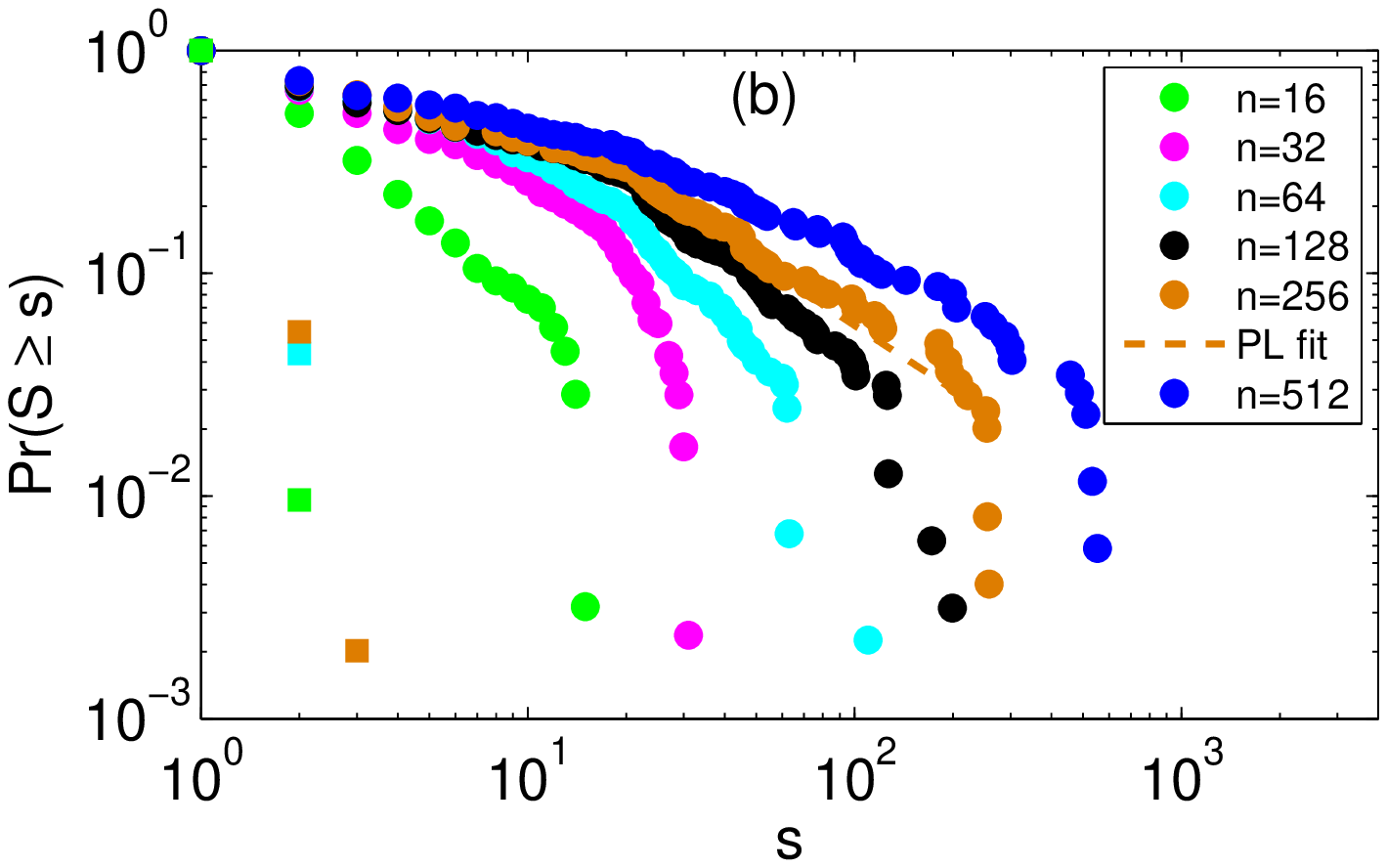}}
  \end{center}
  \caption{Distributions of clusters of significant (a) $H_{xx}$ and (b) $H_{xxx}$ windows of size $s$, for the CAD/EUR pair for different values of the window length $n$. The power-law (PL) fit is done for $n=256$, using the maximum-likelihood method~\cite{cla}. The filled squares correspond to a simulated random time series $\sim N(0,1)$ and different window lengths.}
  \label{fig:log-log}
\end{figure}

\begin{figure}[!t]
  \begin{center}
  \subfigure{\label{fig:log-log_cor_n64}
    \includegraphics[scale=.49]{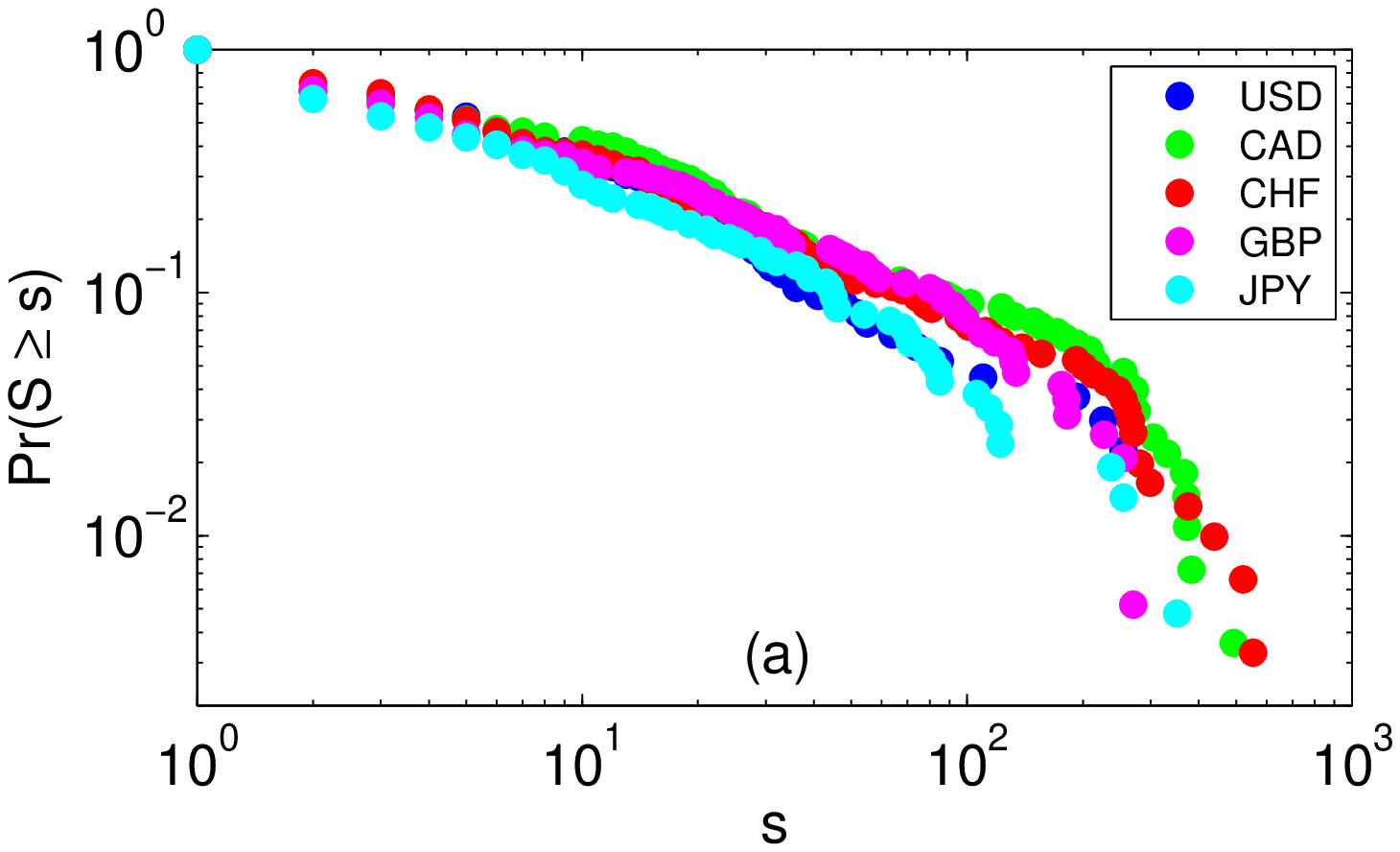}}
      \subfigure{\label{fig:log-log_bic_n64}
    \includegraphics[scale=.49]{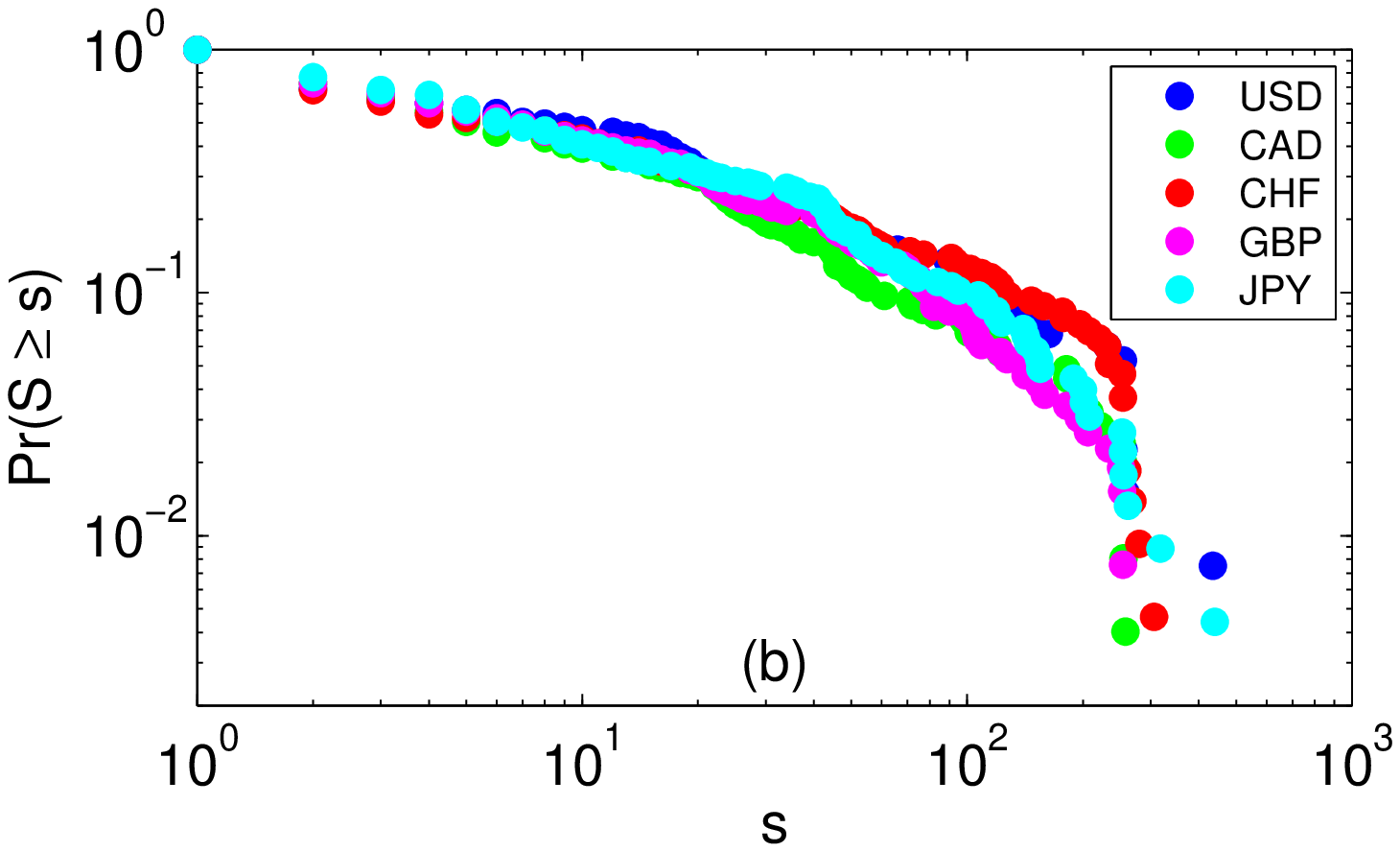}}
  \end{center}
  \caption{Distributions of significant (a) $H_{xx}$ and (b) $H_{xxx}$ cluster sizes $s$ for different currency pairs for the window length $n=256$.}
  \label{fig:log-log_256}
\end{figure}

The epochs of the relative inefficiency represent clusters of instants with significant correlations of various sizes. As pointed out in~\cite{ser,bro1}, the dependency structures are short-lived and not persistent enough to be properly exploited for prediction purposes. Therefore, in this perspective it is interesting to look into the size distribution of the clusters with significant correlations. In Fig.~\ref{fig:log-log} cumulative distributions are plotted in a log-log scale for the CAD/EUR pair with different values of the window length $n$. We can see qualitatively similar behavior in both the significant $H_{xx}$ and $H_{xxx}$ cluster size distributions. Namely, for sufficiently large $n$, fat tails can be observed. As evidenced from Table~\ref{tab:PL_fit}, a reasonable fit\footnote{Normally $p$-value indicates that the null hypothesis is unlikely to be correct and therefore low values are considered good. In~\cite{cla}, by contrast, the $p$-value is used as a measure of the hypothesis that is being verified, and hence high values are good. The power law is a plausible hypothesis for the data if the resulting $p$-value is greater than 0.1. The number of semi-parametric bootstrap repetitions of the fitting procedure is set to 1000.} to the power-law behavior, with the estimated exponent $\hat{\alpha}\approx 1.8$ for $H_{xx}$ and $\hat{\alpha}\approx 1.95$ for $H_{xxx}$, can be obtained by the maximum likelihood method~\cite{cla} for $n=256$. This behavior is in clear contrast with the cluster-size distributions of Gaussian white noise, depicted in Fig.~\ref{fig:log-log} by the square symbols. Similar distributions can also be observed using the remaining currency pairs, as shown in Fig.~\ref{fig:log-log_256}.

\begin{table}
	\centering
		\begin{tabular}{lccccccc}
	  \hline
    & \multicolumn{3}{c}{Significant $H_{xx}$ windows} & \multicolumn{1}{c}{} & \multicolumn{3}{c}{Significant $H_{xxx}$ windows} \\
		\hline \hline
  $n$  & $\hat{x}_{min}$ & $\hat{\alpha}$ & $p$-value & & $\hat{x}_{min}$ & $\hat{\alpha}$ & $p$-value\\
\hline
			$64$ & 4 & 1.77 & 0.00 & & 19 & 2.82 & 0.01 \\
			$128$ & 13 & 1.97 & 0.03  & & 19 &  2.31 & 0.08\\
			$256$ & 12 & 1.8  & {\bf{0.12}} & & 17 & 1.95 & {\bf{0.14}}\\
			\hline
		\end{tabular}
	\caption{Power-law fits to distributions of clusters of significant $H_{xx}$ and $H_{xxx}$ windows for the CAD/EUR pair for different window lengths $n$, using the maximum-likelihood fitting method with goodness-of-fit tests based on the Kolmogorov-Smirnov statistic~\cite{cla}. The parameters $\hat{x}_{min}$ and $\hat{\alpha}$ correspond to the estimates of the lower-bound of the power-law behavior and the scaling exponent, respectively. Statistically significant values with $p>0.1$ are in bold.}
	\label{tab:PL_fit}
\end{table}

\subsection{Predictability in significant $H_{xx}$ windows}
\begin{figure}[!t]
  \begin{center}
  \subfigure[]{\label{fig:hr_pair2_n256}
    \includegraphics[scale=.45]{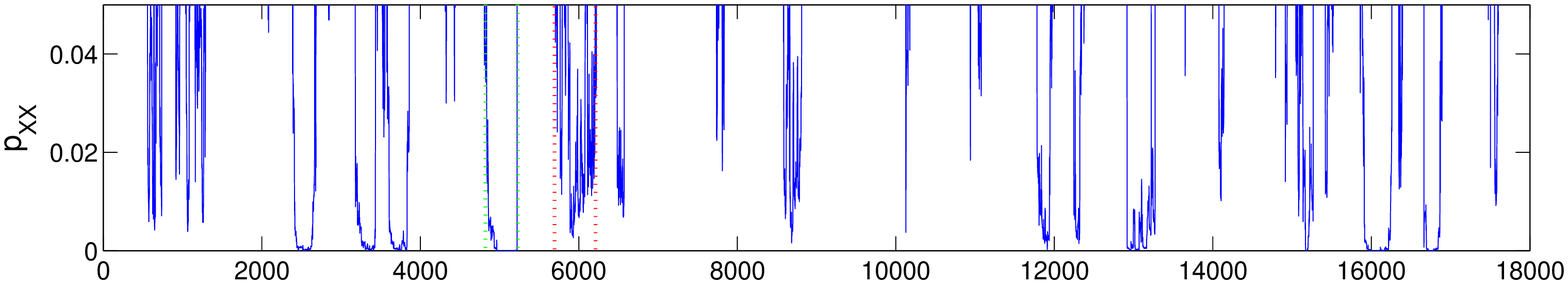}}
    \subfigure[]{\label{fig:Hxx_pair2_n256}
    \includegraphics[scale=.45]{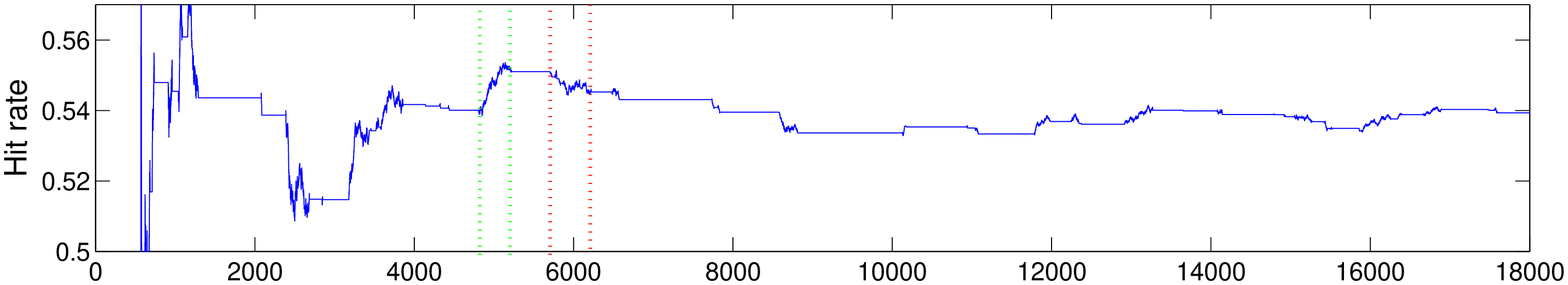}}
  \end{center}
  \caption{Evolution of (a) the $p$-value of the $H_{xx}$ statistic ($p_{xx}$) and (b) the hit rate in the periods of significant correlations. The data correspond to the first 18,000 CAN/EUR return samples and the window size $n=256.$}
  \label{fig:pair2_n256a}
\end{figure}

\begin{figure}[!ht]
  \begin{center}
  \subfigure{\label{fig:hr_wind_pair2}
    \includegraphics[scale=.49]{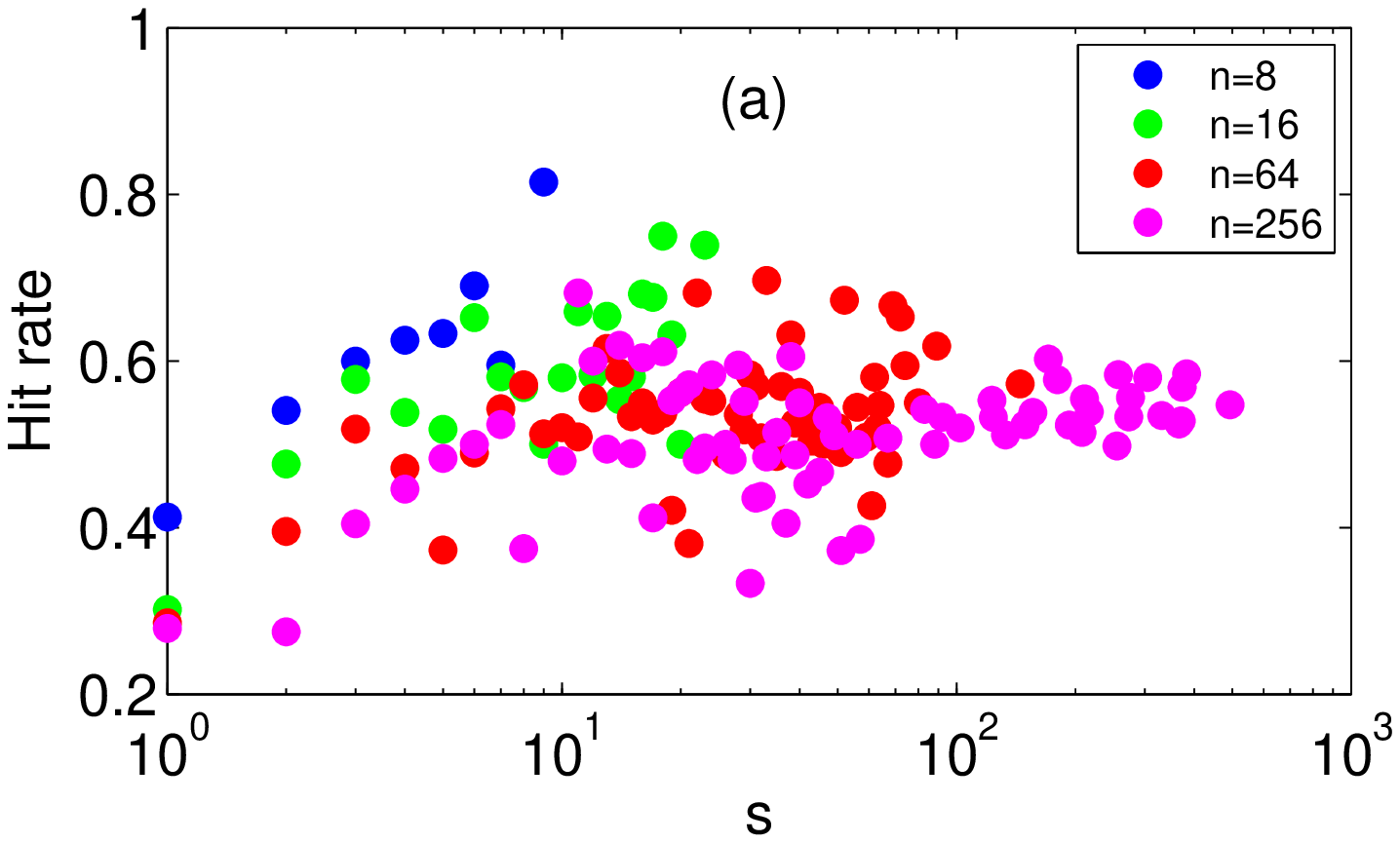}}
      \subfigure{\label{fig:hr_wind_randa}
    \includegraphics[scale=.49]{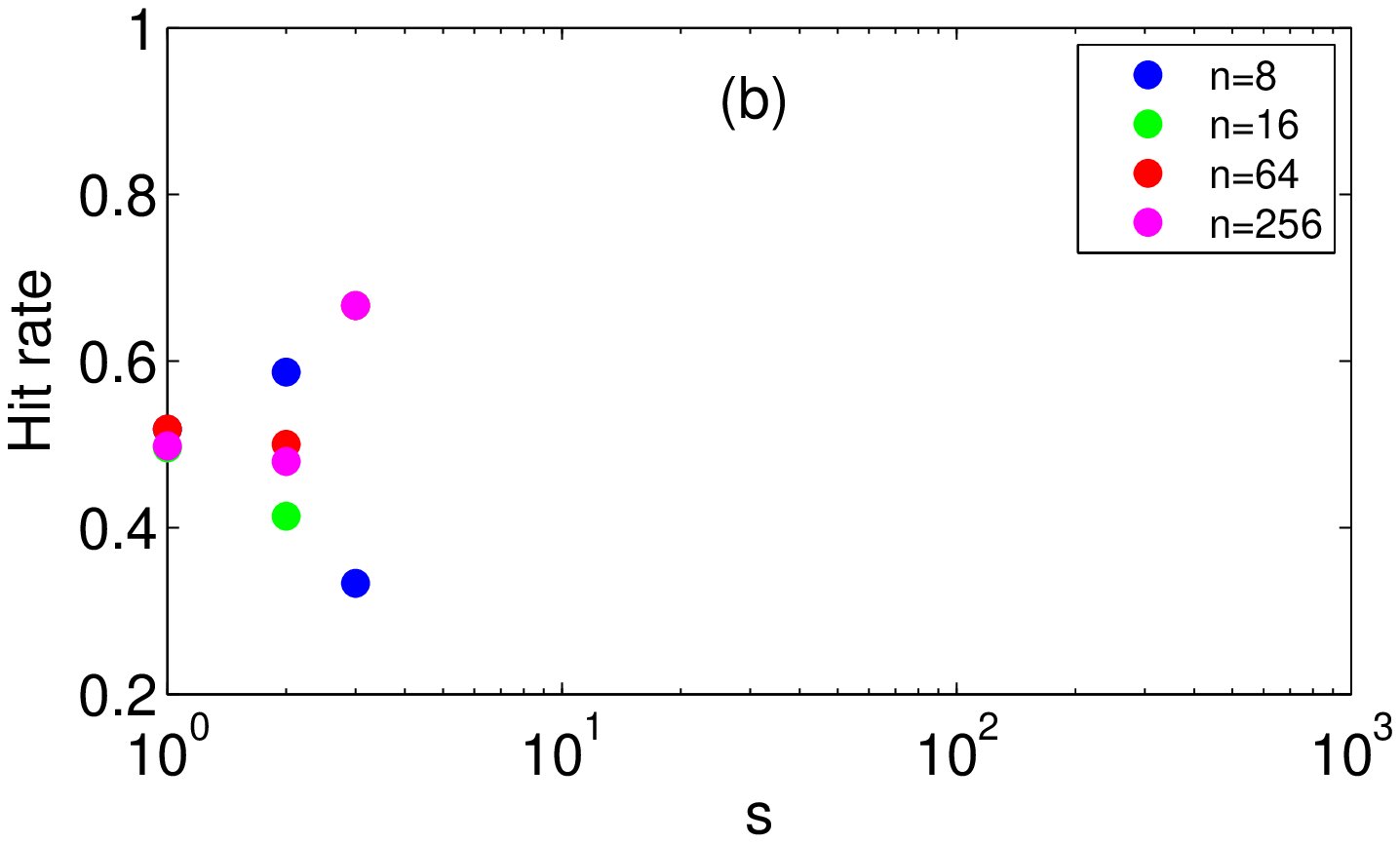}}
  \end{center}
  \caption{Prediction hit rate as a function of the cluster size $s$ for (a) the CAD/EUR pair and (b) a simulated random time series $\sim N(0,1)$, using different window lengths $n$.}
  \label{fig:hr_wind}
\end{figure}

The epochs with significant correlations can be used to forecast the next price change by using the correlation-based predictor (\ref{predict}). Its performance in relation with the significance of the linear correlations is demonstrated in Fig.~\ref{fig:pair2_n256a}, for a selected period of CAD/EUR data. The cumulative hit rate displays a sharp increase during the epochs characterized by persistent contiguous clusters with very low $p$-values of the $H_{xx}$ statistic (e.g., the epoch within the green bars). The hit rate typically deteriorates at the end of these epochs due to the fact that, especially for larger window lengths $n$, the response to changes in the correlation strength is relatively slow and the low $p$-values persist for some time even after the transient dependences have disappeared. A quicker response can be achieved using smaller window length $n$, but then the epochs with transient dependences tend to become shorter, as already evidenced in Fig.~\ref{fig:log-log}. On the other hand, the predictions made in the epochs with $p$-values below the threshold $\alpha=0.05$ but with higher variability are not as successful (e.g., the epoch within the red bars). The tendency of obtaining a better hit rate as the cluster size increases is demonstrated in Fig.~\ref{fig:hr_wind_pair2} for the CAD/EUR pair and different values of the window size $n$. An excellent hit rate is achieved in relatively large clusters for smaller values, such as $n=8,16$. However, as shown in Fig.~\ref{fig:log-log}, such clusters are scarce and the overall hit rate is governed by the dominant small clusters that show poorer hit rate. For larger values of $n$ (e.g., $n=256$), the fat tails tell us that large clusters are more ubiquitous, nevertheless, the hit rate in these large clusters is markedly worse than in the those produced by small $n$. As a result, looking at the total hit rate, no significant differences are observed between different values of the window size $n$. In Fig.~\ref{fig:hr_wind_randa}, we again show qualitative difference compared to the uncorrelated random time series, which does not not feature larger clusters with significant correlations and the hit rate is not affected by the cluster size but randomly fluctuates around the value of 0.5.

\section{Conclusions}
\label{conclusions}

The objective of this paper was to study the dynamics of the linear and non-linear serial dependencies in financial time series, using a windowed portmanteau test procedure in a rolling window framework. In particular, we focused on detection of episodes of statistically significant two- and three-point correlations in returns of several leading currency exchange rates that could offer some potential for their predictability. Since the correlations, which can be viewed as an indicator of market efficiency, are not static but they evolve in time, we employed a rolling window approach in order to capture their dynamics for different window lengths. By analyzing the distributions of the periods with statistically significant correlations it was found that the percentage of the periods with statistically significant two- and three-point correlations in a window of past $n$ observations varies with $n$ and different currencies show different degrees of the dependences. For example, the CHF/EUR pair shows a relatively high percentage of significant correlations, which increases with the window length, while the USD/EUR pair displays a relatively small percentage with little sensitivity to the window length. We found that for sufficiently large window lengths these distributions of the periods with statistically significant correlations fit well to power-law behavior. We also measured the predictability itself by a hit rate, i.e. the rate of consistency between the signs of the actual returns and their predictions, obtained from a simple correlation-based predictor. It was found that during these relatively brief periods the returns are predictable to a certain degree and the presence of large contiguous clusters of instants with significant correlations increase potential for predictability. 

In the present study we did not attempt to use the detected non-linearities in the windows with significant bicorrelations for prediction purposes. Prediction of such non-linear processes is a challenging task, nevertheless, some attempt using a simple model has already been made~\cite{sor}. In a view of the above, it would be interesting to employ jointly linear and non-linear prediction models during the periods featuring solely linear (non-zero correlation and zero bicorrelation) and non-linear (zero correlation and non-zero bicorrelation) dependencies, respectively.

\end{document}